\documentclass[traditabstract]{aa}

\usepackage{graphicx}
\usepackage{txfonts}
\usepackage{float}
\usepackage{mathabx}

\usepackage{natbib}
\bibpunct{(}{)}{;}{a}{}{,}

\usepackage{hyperref}
\hypersetup{colorlinks=true,
            citecolor=blue,
            linkcolor=blue}
\usepackage{color}

\def\vv#1{\vec{#1}}
\def\be {\begin{equation}}
\def\ee {\end{equation}}
\def\vf {\vv{f}}
\def\vg {\vv{g}}
\def\vh {\vv{h}}
\def\vs {\vv{\hat L}}
\def\vn {\vv{\hat G}}
\def\vp {\vv{\hat H}}
\def\vq {\vv{q}}
\def\vu {\vv{u}}
\def\dvq {\vv{\dot q}}
\def\vr {\vv{r}}
\def\vrkl {\vv{r}_{kl}}
\def\ur {\vv{\hat r}}
\def\C{\alpha}
\def\vF{\vv{F}}
\def\vT {\vv{T}}
\def\dvr{\vv{\dot r}}
\def\ddvr{\vv{\ddot r}}
\def\w{\Omega}
\def\vw {\vv{\w}}
\def\tvw {\vv{\tilde \w}}
\def\vL {\vv{L}}
\def\vG {\vv{G}}
\def\vH {\vv{H}}
\def\dvL{\vv{\dot L}}
\def\ia{i}
\def\jb{j}
\def\kc{k}
\def\vi {\vv{\ia}}
\def\vj {\vv{\jb}}
\def\vk {\vv{\kc}}
\def\xii {\hat x}
\def\xj {\hat y}
\def\xk {\hat z}
\def\TI {{\bf \cal I}}
\def\TJ {{\bf \cal J}}
\def\SR {{\bf \cal S}}
\def\M {M}
\def\m {m}
\def\obs{\mathrm{p}}
\def\bin{{}}
\def\sun{s}
\def\ke{k_2}
\def\kf{k_\mathrm{f}}
\def\crm{\cr\noalign{\medskip}}
\def \ij {k}

\def\figpath{}
\def\bibpath{}
\def \llabel#1{\label{#1}}
\def\bfx#1{#1}

\begin{document}

\title{Tidal evolution of the Pluto--Charon binary}

\titlerunning{Tidal evolution of the Pluto--Charon binary}

\author{
Alexandre C. M. Correia\inst{1,2}
}

\authorrunning{A.C.M. Correia}


\institute{
CFisUC, Department of Physics, University of Coimbra, 3004-516 Coimbra, Portugal
\and 
ASD, IMCCE, Observatoire de Paris, PSL Universit\'e, 77 Av. Denfert-Rochereau, 75014 Paris, France
}

\date{\today; Received; accepted To be inserted later}

 \abstract{
A giant collision is believed to be at the origin of the Pluto--Charon system. 
As a result, the initial orbit and spins after impact may have substantially differed from those observed today. More precisely, the distance at periapse may have been shorter, subsequently expanding to its current separation by tides raised simultaneously on the two bodies. Here we provide a general 3D model to study the tidal evolution of a binary composed of two triaxial bodies orbiting a central star. 
We apply this model to the Pluto--Charon binary, and notice some interesting constraints on the initial system. 
We observe that when the eccentricity evolves to high values, the presence of the Sun prevents Charon from escaping because of Lidov-Kozai cycles. 
However, for a high initial obliquity for Pluto or a spin-orbit capture of Charon's rotation, the binary eccentricity is damped very efficiently. 
As a result, the system can maintain a moderate eccentricity throughout its evolution, even for strong tidal dissipation on Pluto.
 } 

 \keywords{
planets and satellites: dynamical evolution and stability ---
minor planets, asteroids: individual (Pluto, Charon)}

 \maketitle


\section{Introduction}

In 1978, a regular series of astrometric observations of Pluto revealed that the images were consistently elongated, denouncing the presence of Pluto's moon, Charon \citep{Christy_Harrington_1978}.
The orbital parameters determined for this system show that the two bodies evolve in an almost circular orbit with a 6.387-day period, and that the system also shows an important inclination of about $122^\circ$ with respect to the orbital plane of Pluto around the Sun \citep[e.g.,][]{Stern_etal_2018}.
Charon has an important fraction of the mass of the system (about 12\%), and therefore can be considered more as a binary planet rather than a satellite.
Indeed, the barycenter of the Pluto--Charon system lies outside the surface of Pluto.
Later, it was found that four additional tiny satellites  move around the barycenter of the system, also in nearly circular and coplanar orbits \citep{Weaver_etal_2006, Brozovic_etal_2015}.

The brightness of Pluto varies by some tens of percent with a period of 6.387~days \citep[e.g.,][]{Walker_Hardie_1955, Tholen_Tedesco_1994, Buie_etal_2010b}.
Although this period coincides with the orbital period of Charon, it has been identified as the rotation of Pluto, since Charon itself is too dim to account for the amplitude of the variation.
Therefore, at present, the rotation of Pluto is synchronous with the orbit of Charon, keeping the same face toward its satellite.
\bfx{The present configuration likely resulted from the action of tidal torques raised on Pluto by Charon.
Tidal torques raised on Charon by Pluto are even stronger, and so the satellite is also assumed to be synchronous with Pluto}, which corresponds to a final equilibrium situation \citep[e.g.,][]{Farinella_etal_1979, Cheng_etal_2014a}.

From photometric observations, \citet{Andersson_Fix_1973} found the angle between the spin of Pluto and its orbit around the Sun to be approximately $90^\circ \pm 40^\circ$.
The uncertainty on this value is significant, but it clearly suggests a high obliquity.
Because of the complete tidal evolution that is evident in the system, the obliquity has usually been assumed to be the same as the inclination of the orbital plane of Charon, that is, $122^\circ$.
Indeed, maps of the surface have been created using HST images and ``mutual events'' \citep[e.g.,][]{Drish_etal_1995, Stern_etal_1997, Young_etal_1999, Buie_etal_2010b}, and although the authors assumed the above obliquity in the creation of these maps, the solution would not have held together if the obliquity was completely incorrect.

Assuming equal densities, the normalized angular momentum density of the Pluto--Charon pair is 0.45 \citep{McKinnon_1989}, exceeding the critical value 0.39, above which no stable rotating single object exists \citep[e.g.,][]{Lin_1981, Durisen_Tohline_1985}.
The proto-planetary disk is not expected to produce such systems, and so alternative theories have been proposed for their origin.
\citet{Harrington_vanFlandern_1979} first suggested that Pluto and Charon could be escaped satellites of Neptune after an encounter with another planet, an unlikely scenario because Triton is on a retrograde orbit \citep{McKinnon_1984}. 
More reliable hypotheses were proposed that take into account the excess of angular momentum in the system, such as binary fission of a rapidly rotating body \citep[e.g.,][]{Lin_1981, Nesvorny_etal_2010} or the accumulation process of planetesimals in heliocentric orbits \citep[e.g.,][]{Tancredi_Fernandez_1991, Schlichting_Sari_2008}.

The above-mentioned theories have some limitations and it is more commonly accepted that Charon resulted from the giant collision of two proto-planets in the early inner Kuiper belt \citep[e.g.,][]{McKinnon_1989, Canup_2005, Rozner_etal_2020}.
This scenario provides the system with its large angular momentum and can also explain the additional small moons in the system \citep{Canup_2011}.
The outward migration of Neptune may have instigated huge perturbations in previously stable zones of the Kuiper belt, and oblique low-velocity collisions between similarly sized objects should have been frequent at the time \citep[e.g.,][]{Malhotra_1993}.
Such a collision probably produced an intact Charon, although it is also possible that a disk of debris orbited Pluto from which Charon later accumulated.
The resulting system is a close binary in an eccentric orbit, with the separation at periapse not exceeding many Pluto radii \citep{Canup_2005, Canup_2011}.

Most previous studies on the past orbital evolution of the Pluto--Charon system \citep{Farinella_etal_1979, Lin_1981,Mignard_1981p, Dobrovolskis_etal_1997, Cheng_etal_2014a}  
assume that both spin axes are normal to the binary orbital plane (2D model), and therefore limit the evolution to the rotations. 
Although this is the expected outcome of tidal evolution, after a large collision the obliquity of Pluto can take any value \citep[e.g.,][]{Dones_Tremaine_1993, Kokubo_Ida_2007, Canup_2011}.
Previous studies also assume that the Pluto--Charon binary is alone.
However, the Sun exerts a torque on the system that causes both the obliquities and the orbital plane of the binary to precess \citep{Dobrovolskis_Harris_1983}.
For the Earth--Moon system, \citet{Touma_Wisdom_1994, Touma_Wisdom_1998} showed that the presence of the Sun is critical to understand its early evolution. 
Some preliminary work on the Pluto--Charon system \citep{Carvalho_2016} suggests that the obliquity and the Sun may also play a role.
Finally, \cite{Cheng_etal_2014a} showed that the inclusion of the gravitational harmonic coefficient $C_{22}$ in the analysis allows smooth, self-consistent evolution to the synchronous state.
It is therefore important to simultaneously take into account the effect of the obliquity, the Sun, and the residual $C_{22}$, in order to obtain a more realistic description of the past history of the Pluto--Charon system.

In Sect.\,\ref{model}, we first derive a full 3D model (for the orbits and spins) that is suitable to describe the tidal evolution of a hierarchical three-body system, where the inner two bodies are assumed to be triaxial ellipsoids. 
In Sect.\,\ref{initcond}, we determine the initial parameters of the Pluto--Charon system that are coherent with the present observations.
In Sect.\,\ref{numsims}, we perform numerical simulations to study the complete evolution of the Pluto--Charon binary. 
Finally, we discuss our results in Sect.\,\ref{concdisc}.

\section{Dynamical model}

\llabel{model}

In this section, we derive a very general model that is suited to study the system composed of Pluto, Charon, and the Sun. 
Pluto and Charon are considered as triaxial ellipsoidal bodies, while the Sun is considered a point-mass (see Fig.\,\ref{fig1}).
Our model is valid in 3D for both orbital planes and individual spins. 
We use Jacobi cartesian coordinates for the orbits, angular momentum vectors for the spins, and quaternions for the rotations.

\begin{figure}
\includegraphics[width=\columnwidth]{\figpath 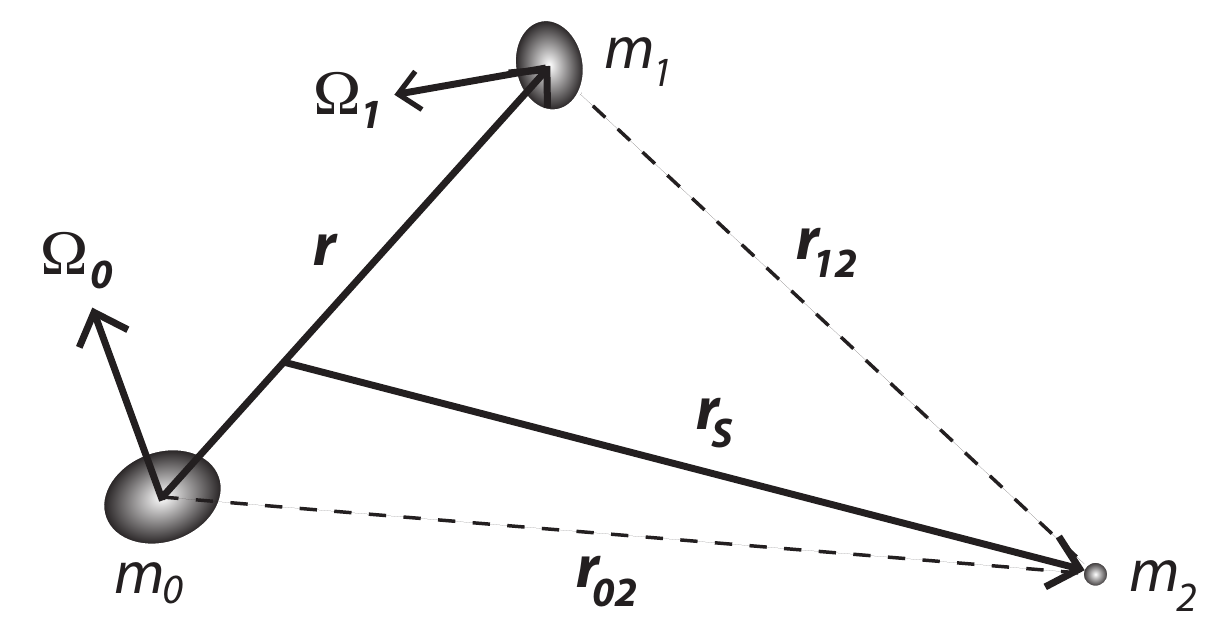}
\caption{Jacobi coordinates, where $\vr_\bin $ is the position of $\m_1$ relative to $\m_0$ (inner orbit), and $ \vr_\sun $ is the position of $\m_2$ relative to the center of mass of $\m_0$ and $\m_1$ (outer orbit). The bodies with masses $\m_0$ and $\m_1$ are considered oblate ellipsoids with angular velocity $\vw$, and body $\m_2$ is considered a point-mass. \llabel{fig1}  }
\end{figure}

\llabel{secmodel}

\subsection{Potential of an ellipsoidal body}

We consider an ellipsoidal body of mass $\m$, and have chosen the cartesian inertial frame ($\vi,\vj,\vk$) as reference.
In this frame, the angular velocity and the rotational angular momentum vectors of the body are given by $\vw = (\w_\ia, \w_\jb, \w_\kc)$ and $\vL = (L_\ia, L_\jb, L_\kc)$, respectively, which are related through the inertia tensor $\TI$ as
\be
\vL = \TI \cdot \vw  \quad \Leftrightarrow \quad \vw = \TI^{-1} \cdot \vL 
\ ,
 \llabel{151019b}
\ee
where
\be
\TI = 
\left[\begin{array}{rrr} 
I_{11}&  I_{12}& I_{13} \crm
I_{12}&  I_{22}& I_{23} \crm
I_{13}&  I_{23}& I_{33} 
\end{array}\right] \ ,
\label{121026c}
\ee
\be\TI^{-1}  = \frac{1}{\Delta \TI}
\left[\begin{array}{ccc} 
I_{22} I_{33} - I_{23}^2   &  I_{13} I_{23} - I_{12} I_{33}   & I_{12} I_{23} - I_{22} I_{13} \crm
I_{13} I_{23} - I_{12} I_{33}   & I_{11} I_{33} - I_{13}^2    & I_{12} I_{13} - I_{11} I_{23} \crm
I_{12} I_{23} - I_{22} I_{13}   & I_{12} I_{13} - I_{11} I_{23}   & I_{11} I_{22} - I_{12}^2
\end{array}\right] \ ,
\label{151028g}
\ee
and
\be
\Delta \TI = I_{11} I_{22} I_{33} + 2 I_{12} I_{13} I_{23} - I_{11} I_{23}^2 - I_{22} I_{13}^2 - I_{33} I_{12}^2 
\ . \label{151028h}
\ee

The gravitational potential of the ellipsoidal body at a generic position $\vr$ relative to its center of mass is given by \citep[e.g.,][]{Goldstein_1950}
\be 
V (\m, \TI, \vr) = - \frac{G \m}{r} + \frac{3 G}{2 r^3} \left[ \ur \cdot \TI \cdot \ur - \frac{1}{3} \mathrm{tr}(\TI) \right] \ , \llabel{121026b}
\ee
where $G$ is the gravitational constant, $\ur = \vr / r = (\xii, \xj, \xk)$ is the unit vector, and $\mathrm{tr}(\TI) = I_{11} + I_{22} + I_{33} $.
We neglect terms in $(R/r)^3$, where $R$ is the mean radius of the body (quadrupolar approximation).
Adopting the Legendre polynomial $P_2(x)=(3x^2-1)/2$, we can rewrite the previous potential as
\begin{eqnarray}
V (\m, \TI, \vr) = - \frac{G \m}{r} 
+ \frac{G}{r^3}  \!\!\!\!\!\! &  \Big[ & \!\!\!\!\!\! \big(I_{22}-I_{11}\big) P_2 (\xj) + \big(I_{33}-I_{11}\big) P_2 (\xk)  \nonumber \\ 
\!\!\!\!\!\! & & \!\!\!\!\!\! + \, 3 \big( I_{12} \xii \xj + I_{13} \xii \xk + I_{23} \xj \xk \big) \Big]
\ . \llabel{151028a}
\end{eqnarray}

\subsection{Point-mass problem}

\llabel{pmp}

We now consider that the ellipsoidal body orbits a point-mass $\M$ located at $\vr$.
The force between the two bodies is easily obtained from the potential energy of the system $ U (\vr) = \M V (\vr) $ as 
\be
\vF = - \nabla U (\vr) =  \vf(\M, \m, \vr) + \vg(\M, \TI, \vr) + \vh(\M, \TI, \vr) 
\ , \llabel{170911d}
\ee
with
\be
\vf(\M, \m, \vr) = - \frac{G \M \m}{r^3} \vr \ ,  \llabel{170911a}
\ee
\begin{eqnarray}
\vg(\M, \TI, \vr) \!\!\!\!\! & = & \!\!\!\!\!  
\frac{15 G \M}{r^5}  \Big[ \frac{I_{22}-I_{11}}{2} \big(\xj^2 - \frac15 \big)  
+ \frac{I_{33}-I_{11}}{2} \big(\xk^2 - \frac15 \big) \nonumber \\ 
&& \!\!\!\!\! + I_{12} \xii \xj + I_{13} \xii \xk + I_{23} \xj \xk \Big] \vr 
\ , \llabel{170911b}
\end{eqnarray}
\begin{eqnarray}
\vh(\M, \TI, \vr) \!\!\!\!\! & = & \!\!\!\!\! 
 -  \frac{3 G \M}{r^4}  \Big[ \big(I_{22}-I_{11}\big) \xj \vj + \big(I_{33}-I_{11}\big) \xk \vk \nonumber \\ 
 && \!\!\!\!\! + I_{12} (\xii \vj + \xj \vi)  + I_{13} (\xii \vk + \xk \vi) + I_{23} (\xj \vk + \xk \vj) \Big]
\ . \llabel{170911c}
\end{eqnarray}
For the orbital evolution of the system, we thus obtain
\be
\ddvr =  \vv{F} / \beta \ ,  \llabel{151028c}
\ee
where $\beta = \M \m/ (\M + \m)$ is the reduced mass.
The spin evolution of the ellipsoidal body can also be obtained from the force by computing the gravitational torque. In the inertial frame we have
\be
\dvL = \vT (\M, \TI, \vr) = - \vr \times \vv{F} = - \vr \times \vh 
\ , \llabel{150626a}
\ee
that is,
\begin{eqnarray}
\vT (\M, \TI, \vr) = \frac{3 G \M}{r^3} \ur \times \Big[ \big(I_{22}-I_{11}\big) \xj \vj + \big(I_{33}-I_{11}\big) \xk \vk && \nonumber \\ 
+ I_{12} (\xii \vj + \xj \vi) + I_{13} (\xii \vk + \xk \vi) + I_{23} (\xj \vk + \xk \vj) \!\!\!\!\!\! &  \Big] & \!\!\!\!\!\!
\ , \llabel{151028d}
\end{eqnarray}
or
\be
\vT =  \frac{3 G \M}{r^3}
\left[
\begin{array}{c}  
\big(I_{33}-I_{22}\big) \xj \xk - I_{12} \xii \xk + I_{13} \xii \xj  + I_{23} (\xj^2 - \xk^2)  \crm 
\big(I_{11}-I_{33}\big) \xii \xk + I_{12} \xj \xk + I_{13} (\xk^2 - \xii^2) - I_{23} \xii \xj  \crm 
\big(I_{22}-I_{11}\big) \xii \xj + I_{12} (\xii^2 - \xj^2) - I_{13} \xj \xk + I_{23} \xii \xk
\end{array}\right] 
\ . \llabel{151028e}
\ee

Apart from a sphere, in the inertial frame ($\vi,\vj,\vk$) the inertia tensor (\ref{121026c}) is not constant.
We let $\SR$ be the rotation matrix, which allows us to convert any vector $\vu_B$ in a frame attached to the body into the cartesian inertial frame $\vu_I$, such that $\vu_I = \SR \, \vu_B$.
Thus, we have
\be
\TI = \SR \, \TI_B \SR^T + \delta \TI   \ ,
\llabel{171124a}
\ee
where $\TI_B = \mathrm{diag}(A, B, C)$ is the permanent deformation inertia tensor in the body frame (expressed in principal axis of inertia), and $\delta \TI$ corresponds to the deformation due to the centrifugal and tidal potentials.
The equilibrium values for each coefficient of $\delta \TI$ are given by \citep{Correia_Rodriguez_2013}:
\be
\frac{\delta I_{11}}{\m R^2} =  \frac{\kf R^3}{3 G \m} \left( \w_\ia^2 -\frac{\w^2}{3} \right) - \ke \frac{\M}{\m} \left(\frac{R}{r}\right)^3 \left( \xii^2 - \frac{1}{3} \right)
 \ , \llabel{151106a}
\ee
\be
\frac{\delta I_{22}}{\m R^2} = \frac{\kf R^3}{3 G \m} \left( \w_\jb^2 -\frac{\w^2}{3} \right) - \ke \frac{\M}{\m} \left(\frac{R}{r}\right)^3 \left( \xj^2 - \frac{1}{3} \right)
\ , \llabel{151106b} 
\ee
\be
\frac{\delta I_{33}}{\m R^2} = \frac{\kf R^3}{3 G \m} \left( \w_\kc^2 -\frac{\w^2}{3} \right) - \ke \frac{\M}{\m} \left(\frac{R}{r}\right)^3 \left( \xk^2 - \frac{1}{3} \right)
\ , \llabel{151106c} 
\ee
\be
\frac{\delta I_{12}}{\m R^2} = \frac{\kf R^3}{3 G \m} \w_\ia \w_\jb - \ke \frac{\M}{\m} \left(\frac{R}{r}\right)^3 \xii \xj
\ , \llabel{151106d} 
\ee
\be
\frac{\delta I_{13}}{\m R^2} = \frac{\kf R^3}{3 G \m} \w_\ia \w_\kc - \ke \frac{\M}{\m} \left(\frac{R}{r}\right)^3 \xii \xk
\ , \llabel{151106e}
\ee
\be
\frac{\delta I_{23}}{\m R^2} = \frac{\kf R^3}{3 G \m} \w_\jb \w_\kc - \ke \frac{\M}{\m} \left(\frac{R}{r}\right)^3 \xj \xk
\ , \llabel{151106f} 
\ee
where $\kf$ and $\ke$ are the fluid and the elastic second Love numbers for potential, respectively (see Sect.\,\ref{tidaldiss} for more details).
The evolution of $\SR$ over time is given by 
\be
\dot \SR = \tvw \, \SR  \ , \quad \mathrm{and} \quad \dot \SR^T = - \SR^T \tvw  
\ , \llabel{160127b}
\ee
with 
\be
\tvw = \left[\begin{array}{ccc} 
0   &  - \w_\kc   & \w_\jb \crm
\w_\kc   & 0   & - \w_\ia \crm
- \w_\jb   & \w_\ia   & 0
\end{array}\right] 
\ . \llabel{160111b}
\ee
In order to simplify the evolution of $\SR$, a set of generalized coordinates can be used to specify the orientation of the two frames.
Euler angles are a common choice, but they introduce some singularities.
Therefore, here we use quaternions \citep[e.g.,][]{Kosenko_1998}.
We denote $\vq = (q_0, q_1, q_2, q_3)$ the quaternion that represents the rotation from the body frame to the inertial frame.
Consequently,
\be
\SR = \left[
\small
\begin{array}{ccc} 
q_0^2+q_1^2-q_2^2-q_3^2   &  2 (q_1 q_2 - q_0 q_3)   & 2 (q_1 q_3 + q_0 q_2) \crm
2 (q_1 q_2 + q_0 q_3)   & q_0^2-q_1^2+q_2^2-q_3^2   & 2 (q_2 q_3 - q_0 q_1) \crm
2 (q_1 q_3 - q_0 q_2)   & 2 (q_2 q_3 + q_0 q_1)   & q_0^2-q_1^2-q_2^2+q_3^2
\end{array}
\normalsize \right] 
\ , \llabel{160127d}
\ee
and
\be
\dvq = \frac{1}{2} (0,\vw) \cdot \vq 
= \frac{1}{2} \left[\begin{array}{c} 
-\w_\ia q_1 - \w_\jb q_2 - \w_\kc q_3 \crm 
\w_\ia q_0 + \w_\jb q_3 - \w_\kc q_2 \crm 
- \w_\ia q_3 + \w_\jb q_0 + \w_\kc q_1 \crm 
\w_\ia q_2 - \w_\jb q_1 + \w_\kc q_0  
\end{array}\right] 
\ . \llabel{160127c}
\ee

To solve the spin-orbit motion, we need to integrate equations (\ref{151028c}), (\ref{150626a}), and  (\ref{160127c}) using the relations (\ref{151019b}), (\ref{171124a}), and (\ref{160127d}).

\subsection{Pluto--Charon binary}
\llabel{pcbin}

Pluto and Charon are considered ellipsoidal bodies with masses $\m_0$ and $\m_1$ and inertia tensors $\TI_0$ and $\TI_1$, respectively, that orbit around each other at a distance $\vr$ from their centers of mass.
The total potential energy can be written from expression (\ref{121026b}) as
\be 
U (\vr) = - \frac{G \m_0 \m_1}{r} + \frac{3 G}{2 r^3} \left[ \ur \cdot \TJ \cdot \ur - \frac{1}{3} \mathrm{tr}(\TJ) \right] \ , \llabel{170911e}
\ee
with $\TJ = \m_0 \TI_1 + \m_1 \TI_0$.
This potential is very similar to the previous point-mass problem and the equations of motion are simply
\be
\ddvr  = \vF_{01} (\vr) / \beta_\bin \ , \llabel{170911f}
\ee
\be
\dvL_0 = \vT_{01} (\vr) \ , \quad  \dvL_1 = \vT_{10} (\vr)
\ , \llabel{170911g}
\ee
\be
\dvq_0 = \frac{1}{2} (0,\vw_0) \cdot \vq_0 \ , \quad \dvq_1 = \frac{1}{2} (0,\vw_1) \cdot \vq_1 \ , \llabel{170911h}
\ee
where
\begin{eqnarray}
\vF_{01} (\vr)  \!\!\!\!\! & = & \!\!\!\!\! 
\vf(\m_0, \m_1, \vr) + \vg(\m_0, \TI_1, \vr) + \vg(\m_1, \TI_0, \vr) \nonumber \\ 
 && \!\!\!\!\! + \vh(\m_0, \TI_1, \vr) + \vh(\m_1, \TI_0, \vr)
\ , \llabel{171127a}
\end{eqnarray}
\be
 \vT_{kl} (\vr) =  \vT (\m_l, \TI_k, \vr)
\ , \llabel{170911g}
\ee
and $\beta_\bin = \m_0 \m_1 / (\m_0 + \m_1)$.
$\vL_k = \TI_k \vw_k $ is the rotational angular momentum vector of the body with mass $\m_k$, $\vw_k$ is the angular velocity vector, and $\vq_k$ is the quaternion that represents the rotation from the body frame to the inertial frame.

\subsection{Effect of the Sun}
\llabel{nbody}

We now consider that the presence of the Sun, with mass $\m_2$, disturbs the Pluto--Charon binary.
We use Jacobi canonical coordinates, which are the distance between the centers of mass of Pluto and Charon,
$\vr_\bin$, and the distance between the center of mass of the binary orbit and the Sun, $ \vr_\sun $ (see Fig.\,\ref{fig1}). 
The total potential energy can be written from expressions (\ref{121026b}) and (\ref{170911e}) as
\be
U_T = U(\vr_\bin) + \m_2 \, \Big[ V(\m_0, \TI_0, \vr_{02}) + V(\m_1, \TI_1, \vr_{12}) \Big]
\ , \llabel{171124b}
\ee
where 
\be
\vr_{02} = \vr_\sun + \frac{\beta_\bin}{\m_0} \vr_\bin
\ , \quad \mathrm{and} \quad 
\vr_{12} = \vr_\sun - \frac{\beta_\bin}{\m_1} \vr_\bin 
\ . \llabel{171127c}
\ee

The equations of motion for the orbits and spins are 
\be
\ddvr_\bin = \frac{1}{\beta_\bin} \vF_{01} (\vr_\bin) + \frac{1}{\m_0} \vF_{02} (\vr_{02}) - \frac{1}{\m_1} \vF_{12} (\vr_{12}) 
\ , \llabel{171127d}
\ee
\be
\ddvr_\sun = \frac{1}{\beta_\sun} \Big[ \vF_{02} (\vr_{02}) + \vF_{12} (\vr_{12}) \Big] 
\ , \llabel{171127e}
\ee
\be
\dvL_0 =  \vT_{01} (\vr_\bin) +  \vT_{02} (\vr_{02}) \ ,  \llabel{171124c}
\ee
\be
\dvL_1 = \vT_{10} (\vr_\bin) + \vT_{12} (\vr_{12})  \ ,  \llabel{171124d}
\ee
\be
\dvq_0 = \frac{1}{2} (0,\vw_0) \cdot \vq_0 \ , \quad \dvq_1 = \frac{1}{2} (0,\vw_1) \cdot \vq_1 
\ , \llabel{170911l}
\ee
where $\beta_\sun = \m_2 (\m_0 + \m_1) / (\m_0 + \m_1 + \m_2)$, $\vF_{01} (\vr_\bin)$ and $\vT_{kl} (\vrkl)$ are given by expressions (\ref{171127a}) and  (\ref{170911g}), respectively, and
\begin{eqnarray}
\vF_{k2} (\vr)  \!\!\!\!\! & = & \!\!\!\!\! 
\vf(\m_k, \m_2, \vr) + \vg(\m_2, \TI_k, \vr) + \vh(\m_2, \TI_k, \vr)
\ . \llabel{171127b}
\end{eqnarray}

\subsection{Tidal evolution}
\llabel{tidevol}

The equations of motion derived in Sect.\,\ref{nbody} conserve the total energy of the system.
They already take into account the tidal bulges (Eqs.\,(\ref{151106a})$-$(\ref{151106f})), but not tidal dissipation.
\bfx{The dissipation of the mechanical energy of tides inside the bodies introduces a time delay $\Delta t$, and hence a phase shift, between the initial perturbation and the maximal deformation.
As a consequence, there is an additional net torque on the tidal bulges, which modify the spins and the orbits.}

Tidal dissipation is usually modeled through the elastic second Love number $\ke$ and the quality factor $Q$.
The first is related to the rigidity of the body and measures the amplitude of the tidal deformation, while the second is related with the viscosity and measures the amount of energy dissipated in a tidal cycle \citep[e.g.,][]{Munk_MacDonald_1960}.
For a given tidal frequency, $\sigma$, the tidal dissipation can be related to this delay through \citep[e.g.,][]{Correia_Laskar_2003JGR}
\be
Q_\sigma^{-1} = \sin ( \sigma \Delta t_\sigma ) \approx \sigma \Delta t_\sigma \ .
\llabel{171013b}
\ee
The exact dependence of $\Delta t_\sigma$ on the tidal frequency is unknown.
In order to take into account tidal dissipation, we need to adopt a tidal model.
A large variety of models exist, but the most commonly used are the constant-$Q$ \citep[e.g.,][]{Munk_MacDonald_1960}, the linear model \citep[e.g.,][]{Mignard_1979}, 
the Maxwell model \citep[e.g.,][]{Correia_etal_2014},
and the Andrade model \citep[e.g.,][]{Efroimsky_2012}.
Some models appear to be better suited to certain situations, but there is no model that is globally accepted.
Nevertheless, regardless of the tidal model adopted, the qualitative conclusions are more or less unaffected, and the system always evolves into a minimum of energy \citep[e.g.,][]{Hut_1980}.

Here we adopt a viscous linear model for tides \citep{Singer_1968, Mignard_1979}.
In this model it is assumed that the time delay is constant and independent of the frequency.
This choice is motivated by the fact that most of the tidal evolution in the Pluto--Charon binary occurs in just a few million years after formation (see Sect.\,\ref{numsims}), when the two bodies are likely still mostly melt and fluid.
Moreover, the linear tidal model provides very simple expressions for the tidal interactions that are valid for any eccentricity, inclination, rotation, and obliquity.

As in Sect.\,\ref{pmp}, we consider an ellipsoidal body with mass $\m$ that orbits a point-mass $\M$ located at $\vr$.
The total tidal force acting on the orbit is given by \citep[e.g.,][]{Mignard_1979}
\be
\vF_t (K, \M, \vw, \vr) = - K \frac{\M^2}{r^{10}} \Big[ 2 (\vr \cdot \dvr ) \vr + r^2 (\vr \times \vw + \dvr ) \Big] \ , 
\llabel{171016a}
\ee
and the tidal torque on the spin
\be
\vT_t (K, \M, \vw, \vr) = - \vr \times \vF_t = K \frac{\M^2}{r^8} \Big[ (\vr \cdot \vw) \vr - r^2\vw + \vr \times \dvr \Big]  \ ,
\llabel{171016c}
\ee
where
\be
K =  3 G R^5 \ke \Delta t
\llabel{171016b}
\ee
contains all the quantities pertaining to the body with mass $\m$.
We can now add to the equations (\ref{171127d})$-$(\ref{171124d}), the contribution of the tidal evolution of the Pluto--Charon system as
\begin{eqnarray}
\ddvr_\bin \!\!\!\!\! & = & \!\!\!\!\!  
\frac{1}{\beta_\bin} \Big[ \vF_t (K_0, \m_1, \vw_0, \vr_\bin) + \vF_t (K_1,\m_0, \vw_1, \vr_\bin) \Big]   \nonumber \\  && \!\!\!\!\! 
+ \frac{1}{\m_0} \vF_t (K_0, \m_2, \vw_0, \vr_{02}) - \frac{1}{\m_1} \vF_t (K_1, \m_2, \vw_1, \vr_{12})
\ , \llabel{190906a}
\end{eqnarray}
\be
\ddvr_\sun = \frac{1}{\beta_\sun} \Big[ \vF_t (K_0, \m_2, \vw_0, \vr_{02}) + \vF_t (K_1, \m_2, \vw_1, \vr_{12}) \Big] 
\ , \llabel{190906b}
\ee
\be
\dvL_0 =  \vT_t (K_0, \m_1, \vw_0, \vr_\bin) + \vT_t (K_0, \m_2, \vw_0, \vr_{02})  
 \ ,  \llabel{190906c}
\ee
\be
\dvL_1 =  \vT_t (K_1, \m_0, \vw_1, \vr_\bin) + \vT_t (K_1, \m_2, \vw_1, \vr_{12})  
\ .  \llabel{190906d}
\ee

\section{Initial conditions}
\llabel{initcond}

The commonly accepted scenario for the formation of the Pluto--Charon binary is a giant impact of two proto-planets \citep[][]{McKinnon_1989, Canup_2005, Canup_2011}.
The resulting system is a packed binary in an eccentric orbit, with the separation at periapse not exceeding a few Pluto radii \citep{Canup_2005, Canup_2011}.
In Table~\ref{TabIn} we show three possible examples of initial configurations with different initial eccentricities taken from different works.

  \begin{table}
    \begin{center}
      \caption{Possible initial configurations for the Pluto--Charon orbit.}
      \begin{tabular}{cccc}
        \hline
         orbit & $a / R_0 $ & $e$ & reference \\ \hline
        \#1 & 4.0 & 0.20 & \citet{Cheng_etal_2014a}  \\
        \#2 & 6.5 & 0.50 & \citet{Canup_2005}  \\
        \#3 & 15.8 & 0.77 & \citet{Canup_2011}  \\
        \hline
      \end{tabular}
      \label{TabIn}
    \end{center}
  \end{table}

Although the initial orbits can be quite different, there are some constraints on the system, such that tides can bring it to the present observed configuration \citep[][]{Farinella_etal_1979, Lin_1981, Mignard_1981p, Dobrovolskis_etal_1997, Cheng_etal_2014a}.
We use these constraints to determine the starting point of the numerical simulations in Sect.\,\ref{numsims}.

\subsection{Angular momentum}

If we neglect the effect of the Sun, 
the total angular momentum of the binary, $\vH$, must be conserved.
This property can be used to put constraints on the initial spin states of Pluto and Charon, $\vL_0$ and $\vL_1$, respectively.
We let (Eq.\,(\ref{151019b}))
\be
\vL_k = \TI_k \vw_k = L_k \, \vs_k \approx C_k \w_k \, \vs_k 
\ , \llabel{190909c}
\ee
with $L_k = | \vL_k | $, $\vs_k = \vL_k / L_k $, and
$C_k$ being the moment of inertia with respect to the spin axis.
The binary orbital angular momentum is
\be
\vG_\bin = \beta_\bin \, \vr_\bin \times \dvr_\bin = G \, \vn =
\beta_\bin n_\bin a_\bin^2 \sqrt{1-e_\bin^2} \, \vn  
\ , \llabel{190909a}
\ee
where $a_\bin$ is the semi-major axis, $e_\bin$ is the eccentricity, and $n_\bin$ is the orbital mean motion.
Consequently,
\be
\vH = \vL_0 + \vL_1 + \vG = H \, \vp = n_\obs \, \big( C_0 + C_1 + \beta_\bin a_\obs^2 \big) \, \vp = const 
\ , \llabel{190909b}
\ee
where $a_\obs$ and $n_\obs$ are the presently observed semi-major axis and mean motion, respectively.
We assume that the present spins are aligned with the orbit normal and that both bodies are synchronous, because this corresponds to the last stage of tidal evolution \citep{Hut_1980}.
We additionally denote $\theta_k$ the obliquity, that is, the angle between the spin and the orbit, such that
\be
\cos \theta_k = \vs_k \cdot \vn 
\ , \llabel{190909d}
\ee
and $I$ the inclination between the initial and the present orbit of the binary, such that
\be
\cos I = \vn \cdot \vp
\ . \llabel{190909d}
\ee

The rotational angular momentum of Charon is the smallest contribution in the total angular momentum.
We therefore further assume for simplicity that the initial obliquity of Charon is zero, that is, $\vs_1 = \vn$.
From expression (\ref{190909b}) we have
\be
H^2  = L_0^2 + (L_1 + G)^2 + 2 \, (L_1 + G) \, L_0 \cos \theta_0 
\ , \llabel{190910a}
\ee
and
\be
H \cos I = L_0  \cos \theta_0 + L_1 + G  
\ . \llabel{190910b}
\ee

The above equations give us two constraints for the initial spins, provided that we know the initial orbit (Table~\ref{TabIn}).
In general, $L_1 \ll G$, and so we can completely determine the initial spin of Pluto from the initial orbit, characterized by $\vG$:
\be
L_0 = \sqrt{H^2 + (L_1 + G)^2 - 2 \, (L_1 + G) \, H \cos I} 
\approx | \vH - \vG |
\ , \llabel{190910c}
\ee
and
\be
\cos \theta_0 = \frac{H \cos I - L_1 - G}{L_0} \approx \frac{H \cos I - G}{ | \vH - \vG |}
\ . \llabel{190910d}
\ee

In the formation scenarios (Table~\ref{TabIn}), the inclination between the initial and the present orbit of the binary, $I$, is not provided.
As this parameter is connected with the initial spin of Pluto, in our numerical simulations (Sect.\,\ref{numsims}) we assume different values for the initial obliquity $\theta_0$ and then derive constraints for
\be
L_0 = \sqrt{H^2 - (L_1 + G)^2 \sin^2 \theta_0}  - (L_1 + G) \cos \theta_0
\ , \llabel{200526a}
\ee
and
\be
\cos I = \frac{L_0  \cos \theta_0 + L_1 + G}{H}
\ . \llabel{200526b}
\ee

\subsection{Rotation}
\llabel{rotlimit}

The centrifugal breakup period $\sqrt{3\pi/G\rho}$ of Pluto and Charon is about 2.5~hours, where $\rho$ is the mean density.
We can assume this rotation period as the critical value for the initial rotation immediately after formation.
Orbital solutions that provide a rotational angular momentum (Eq.\,(\ref{190910c})) that is not compatible with this critical value can be excluded from the simulations.

Assuming principal axis rotation, we can obtain the initial rotation rate from the rotational angular momentum as (Eq.\,(\ref{151019b}))
\be
\w_k = L_k / C_k \ . \llabel{190910x}
\ee
For a homogenous sphere we have $C / (\m R^2) = 2/5 $.
Adopting a two-layer body with densities of 3.4~g/cm$^3$ and 0.95~g/cm$^3$ for rock and ice, respectively, we estimate the core radius \citep{Nimmo_etal_2017} and obtain a more realistic value $C / (\m R^2) \approx 0.3$. 
Moreover, for fast-rotating bodies, the centrifugal potential modifies the mass distribution about the spin axis and introduces a correction in the inertia tensor $C = 0.3 m R^2 +\delta C$, with  (Eq.\,(\ref{151106c}))
\be
\frac{\delta C}{\m R^2} = \kf \frac{2 \w^2 R^3}{9 G \m} = \frac{\kf \w^2}{6 \pi G \rho}
\ , \llabel{190910za} 
\ee
where $\kf$ is the fluid Love number.
\bfx{For a homogeneous body we have $\kf = 3/2$, but for differentiated bodies $\kf$ is always smaller.
Applying the Darwin-Radau relation \citep[e.g.,][]{Jeffreys_1976} we obtain $\kf \approx 0.73$ for the two-layer body.}
Inserting this into expression (\ref{190910za}) we estimate
\be
\frac{\delta C}{\m R^2} \approx P_\mathrm{h}^{-2}
\ , \llabel{190910y} 
\ee
where $P_\mathrm{h}$ is the rotation period in hours.
At present we have $P_\mathrm{h}\approx153$~hours, and so this correction can be neglected. 
\bfx{However, for fast initial rotation periods the correct rotation is obtained by correcting the inertia tensor in the expression of the rotation rate (Eq.\,(\ref{190910x})), and solving the cubic equation}
\be
\w_k \left( 0.3 +  \frac{\kf \w_k^2}{6 \pi G \rho} \right) = \frac{L_k}{\m_k R_k^2} 
\ . \llabel{190910zb}
\ee
We note that the model that we present in Sect.\,\ref{model} already takes into account these corrections, not only for the centrifuge distortion, but also for the less important tidal one (Eq.\,(\ref{151106c})).

For Charon we arbitrarily use 6~hours for the initial rotation period and zero initial obliquity in all our simulations. The initial $L_1$ is directly obtained from expression (\ref{190910zb}).
These values are not critical, because the rotation of Charon quickly evolves into an equilibrium configuration, while its obliquity undergoes large variations (Sect.\,\ref{initobl}).
The initial spin of Pluto is computed from expression (\ref{200526a}), which depends on the initial orbit (see Table~\ref{TabIn}).
The initial $\w_0$ is then obtained by solving equation (\ref{190910zb}).

\subsection{Shape}
\llabel{shapeJC}

The images taken during the {\it New Horizons} spacecraft encounter were used to determine the mean radius and shapes of Pluto and Charon \citep{Nimmo_etal_2017}.
While the radius measurements 
were obtained with good precision (see Table~\ref{TabGeo}), the present-day shapes were inconclusive.
Only upper bounds on the flattening of 0.6\% (7~km) for Pluto and 0.5\% (3~km) for Charon were obtained, 
consistent with hydrostatic equilibrium.
Indeed, from expression (\ref{190910y}) and $\kf = 0.73$ we estimate 
\be
\frac{\delta R}{R} \approx \frac{3}{2} P_\mathrm{h}^{-2}
\ , \llabel{191125a} 
\ee
which yields present-day distortions smaller than 0.1~km, which is too small to be detectable in the {\it New Horizons} images.
The absence of significant deformations for Pluto and Charon implies that their interiors must have been warm and/or deformable during the whole orbital evolution of the system.

Assuming a homogeneous density, we can compute the Stokes' gravity field coefficients from the ellipsoid semi-axes ($ a \ge b \ge c$), such that \citep[e.g.][]{Yoder_1995cnt}: 
\be
J_2 = \frac{1}{5} \frac{a^2+b^2-2c^2}{a^2+b^2} \ , \quad \mathrm{and} \quad C_{22} =  \frac{1}{10} \frac{a^2-b^2}{a^2+b^2} \ .
\llabel{171004b}
\ee
We now assume a distortion $\delta R /R = 10^{-4}$ (i.e., $\delta R \approx 0.1$~km) for Pluto and Charon, which is compatible with the hydrostatic residuals (Eq.\,(\ref{191125a})) and below the observed upper limits \citep{Nimmo_etal_2017}.
Consequently, using $a=R+\delta R$, $b=R$, and $c=R-\delta R$, we obtain $J_2 = 6 \times 10^{-5}$ and $C_{22} = 10^{-5}$.
In our work, we adopt these values as permanent residual deformations for the two bodies.

  \begin{table}
    \begin{center}
      \caption{Geophysical parameters for the Pluto--Charon binary \citep{Nimmo_etal_2017}. 
      $M_\Pluto = 1.30 \times 10^{22}$~kg, and $R_\Pluto = 1.19 \times 10^6$~m .}
      \begin{tabular}{cccc}
        \hline
         param. & unit & Pluto & Charon  \\ \hline
        $\m$ & $M_\Pluto$ & 1.0 & 0.122  \\ 
        $R$ & $R_\Pluto$ & 1.0 & 0.510 \\
        $C$ & $mR^2$ & 0.3 & 0.3  \\
        $J_2$ & $10^{-5}$ & 6.0 & 6.0  \\
        $C_{22}$ & $10^{-5}$ & 1.0 & 1.0  \\
        $\kf$ & $-$ & 0.73 & 0.73  \\
        $\ke \Delta t$ & s & 30   & $15 A$ \\
        \hline
      \end{tabular}
      \label{TabGeo}
    \end{center}
  \end{table}

In the early stages of the system evolution, when the rotations are much faster than today and the two bodies are close to each other, the hydrostatic contribution to $J_2$ and $C_{22}$ is several orders of magnitude above the present residual values.
However, as the system evolves and the bodies cool down, they are expected to freeze at the present deformations.
It is then important to keep some permanent deformation in the bodies, even if extremely small, to lock the rotations at the present synchronous state.

\subsection{Tidal dissipation}
\llabel{tidaldiss}

The elastic second Love number for an incompressible homogeneous body is given by \citep{Love_1911}
\be
\ke = \frac{3}{2} \left(1+\frac{19\mu}{2g\rho R}\right)^{-1} \ ,
\llabel{171013a}
\ee
where $g = G \m / R^2$ is the surface gravity and $\mu$ is the rigidity.
It is common to estimate $\mu \approx 4$\,GPa for icy bodies \citep[e.g.,][]{Nimmo_Schenk_2006}, and so we obtain $\ke \approx 0.05$ for Pluto and $\ke \approx 0.01$ for Charon. 
As tidal dissipation and evolution only depend on the product $\ke \Delta t$ (Eq.\,(\ref{171016b})), we adopt $\ke = 0.05$ for both bodies and then use different $\Delta t$ values for Charon.

Our incomplete knowledge of the physics of tides means that the $\Delta t$ values are unknown.
Yet, since $\Delta t$ is usually small, it only affects the overall timescale of tidal evolution (Eqs.\,(\ref{171016a}) and (\ref{171016c})).
As pointed out by previous studies \citep[e.g.,][]{Ward_Canup_2006, Cheng_etal_2014a}, the ratio between tidal dissipations in Charon and Pluto is the important parameter for the tidal evolution history of the Pluto--Charon binary, namely (Eq.\,(\ref{190906a}))
\be
A \equiv \frac{K_1 m_0^2}{K_0 m_1^2} = \frac{\rho_0^2 R_0 k_{21} \Delta t_1}{\rho_1^2 R_1 k_{20} \Delta t_0} \approx 2 \frac{k_{21} \Delta t_1}{k_{20} \Delta t_0}
\ . \llabel{190911a}
\ee
For Pluto we adopt $\Delta t_0 = 600$~s, the same value as that for the Earth \citep{Dickey_etal_1994,  Touma_Wisdom_1994}, and also the same adopted by a former study of the Pluto--Charon tidal evolution \citep{Cheng_etal_2014a}, for a better comparison.
For Charon we adopt $\Delta t_1 = A \Delta t_0 /2 = 300 \, A$~s, and vary $A$ from 1 to 16.

\subsection{Evolution timescales}
\llabel{evts}

The tidal evolution of the spins and orbits of a binary system perturbed by an external body is given by equations (\ref{190906a}) to (\ref{190906d}).
These equations are general, but in the case of the Pluto--Charon system, the Sun is very distant and its tidal effect can be neglected.
Therefore, we can drop equation (\ref{190906b}) and simplify the remaining ones as ($\ij=0,1$)
\be
\dvL_\ij =  \vT_t (K_\ij, \m_{1-\ij}, \vw_\ij, \vr_\bin) 
 \ ,  \llabel{200530a}
\ee
and
\be
\ddvr_\bin =
\frac{1}{\beta_\bin} \sum_{\ij=0,1} \vF_t (K_\ij, \m_{1-\ij}, \vw_\ij, \vr_\bin)    
\ . \llabel{200530b}
\ee
Moreover, averaging over the mean anomaly and the argument of the pericenter of the orbit, we get simplified versions of these equations in terms of elliptical elements as \citep{Correia_2009}
\be
\frac{\dot \w_\ij}{\w_\ij} = - A_\ij \frac{{\cal K} n}{C_\ij \w_\ij} 
\left( f_1(e) \frac{1+\cos^2 \theta_\ij}{2} \frac{\w_\ij}{n} - f_2(e) \cos \theta_\ij \right) \ , \llabel{090515a}
\ee
\be
\dot \theta_\ij \simeq A_\ij \frac{{\cal K} n}{C_\ij \w_\ij} \sin \theta_\ij
\left( f_1(e) \cos \theta_\ij \frac{\w_\ij}{2 n} - f_2(e) \right) 
\ , \llabel{090520d}
\ee
\be
\frac{\dot a}{a} =  \frac{2 {\cal K}}{\beta a^2} \Big[ 
\Big( f_2(e) X_0 - f_3(e) \Big) 
+ A \Big( f_2(e) X_1 - f_3(e) \Big)  
\Big] \ , \llabel{090515b}
\ee
\be
\frac{\dot e}{e} = \frac{9 {\cal K}}{\beta a^2} \Big[ 
\Big( \frac{11}{18} f_4(e) X_0 - f_5(e) \Big) 
+ A \Big( \frac{11}{18} f_4(e) X_1 - f_5(e) \Big)
  \Big] \ , \llabel{090515c}
\ee
where $
{\cal K} = K_0 \, \m_1^2 / a^6 
$,
$A_\ij = \left( \delta_{0 \ij} + A \, \delta_{1 \ij} \right) $,
$X_\ij = \cos \theta_\ij \, \w_\ij /n$, and $f_k(e)$ 
are functions that depend solely on the eccentricity and become equal to one for $e=0$:
\be f_1(e) = \frac{1 + 3e^2 + \frac38 e^4}{(1-e^2)^{9/2}} \ , \ee
\be f_2(e) = \frac{1 + \frac{15}{2} e^2 + \frac{45}{8} e^4 + \frac{5}{16} e^6}{(1-e^2)^{6}} \ , \ee
\be f_3(e) = \frac{1 + \frac{31}{2} e^2 + \frac{255}{8} e^4 + \frac{185}{16} e^6 + \frac{25}{64}e^8}{(1-e^2)^{15/2}} \ , \ee
\be f_4(e) = \frac{1 + \frac32 e^2 + \frac18 e^4}{(1-e^2)^5} \ , \ee
\be f_5(e) = \frac{1 + \frac{15}{4} e^2 + \frac{15}{8} e^4 + \frac{5}{64} e^6}{(1-e^2)^{13/2}} \ .
\ee

Adopting the current final value of $a_\obs = 16.5 \,R_0$ for the semi-major axis, we estimate the spin evolution timescale of Pluto and Charon, namely,
\be
\tau_0 \sim \frac{C_0}{\cal K} \approx 3 \times 10^5 \, \mathrm{yr} \ ,
\ee
\be
\tau_1 \sim \frac{C_1}{{\cal K} A} \approx \frac{9}{A} \times 10^3 \, \mathrm{yr} \ ,
\ee
respectively, and the orbital evolution timescale
\be
\tau \sim \frac{m_1 a^2}{9 \cal K} \approx 3 \times 10^6 \, \mathrm{yr} \ . \llabel{191215c}
\ee
These quick estimations agree relatively well with what we observe in the numerical simulations, which are extended up to $10^7$~yr to ensure that the system always ends in the present observed state.
We see that we always have $\tau_1 \ll \tau_0  < \tau$, which means that the spins evolve faster than the orbit, and that the spin of Charon evolves much faster than that of Pluto.
The equilibrium rotation is obtained when $\dot \w_\ij = 0$ (Eq.\,(\ref{090515a})) for
\be
\frac{\w_{\rm e}}{n} = \frac{f_2(e)}{f_1(e)} \, \frac{2 \cos
\theta_\ij}{1 + \cos^2 \theta_\ij} 
\ . \llabel{090520a}
\ee

As the spin of Charon evolves much faster than anything else, we can replace $\w_1=\w_{\rm e}$ with $\theta_1=0$ in expressions (\ref{090515b}) and (\ref{090515c}) to get simplified expressions:
\be
\frac{\dot a}{a} =  \frac{2 {\cal K}}{\beta a^2} \Big[
\Big( f_2(e) \cos \theta_0 \frac{\w_0}{n} - f_3(e) \Big) 
- \frac{7 A}{18}  f_6(e) e^2  \Big] \ , \llabel{200602b}
\ee
\be
\frac{\dot e}{e} = \frac{9 {\cal K}}{\beta a^2} \Big[ 
\Big( \frac{11}{18} f_4(e) \cos \theta_0 \frac{\w_0}{n} - f_5(e) \Big) 
- \frac{7 A}{18} f_6(e) (1-e^2)
  \Big] \ ,
\llabel{200602c}
\ee
with
\be f_6(e) = \frac{1 + \frac{45}{14} e^2 + 8 e^4 + \frac{685}{224} e^6 + \frac{255}{448} e^8 + \frac{25}{1792} e^{10}}{(1 + 3e^2 + \frac38 e^4) (1-e^2)^{15/2}} \ . \ee

For large $A$ values, the eccentricity evolution is dominated by tides raised on Charon, and so the orbit is circularized in the early stages of the evolution \citep{Dobrovolskis_etal_1997}.
On the other hand, for small $A$ values, the evolution is dominated by tides raised on Pluto, whose rotation decreases slowly, and therefore the eccentricity is allowed to grow to higher values \citep{Cheng_etal_2014a}.
For intermediate $A$ values, 
the eccentricity may preserve a nonzero but not overly high value for most of the evolution.

\subsection{Orbit of the Sun}
\llabel{sunorb}

Previous constraints were derived assuming a two-body problem.
However, in our numerical simulations, we additionally consider the presence of the Sun (Sect.\,\ref{nbody}). 
The final orbit of the Pluto--Charon binary around the Sun is assumed to be exactly the same orbit as today, that is, it has a semi-major axis $a_\sun = 39.5$~au and an eccentricity $e_\sun = 0.25$.
More importantly, we assume that the inclination between the present orbit of the binary and the orbit of the Sun is $i=122^\circ$ \citep{Stern_etal_2018}.

The semi-major axis and the eccentricity of the Sun remain almost unchanged throughout the evolution of the Pluto--Charon binary, because we do not include the effect from the remaining planets in our study.
However, we note that the initial mutual inclination may change when Pluto has an initial nonzero obliquity (Eq.\,(\ref{200526b})).
We therefore chose the initial mutual inclination between the orbit of the binary and the orbit of the Sun to be $i = (122^\circ + I)$, such that it will stabilize at the present value at the end of the tidal evolution (when $I=0^\circ$).

\section{Numerical simulations}
\llabel{numsims}

In this section we simulate the tidal evolution of the Pluto--Charon binary from the early stages of its formation until the present day configuration.
We numerically integrate equations (\ref{171127d}) to (\ref{170911l}) for the conservative motion of the orbits and spins, and equations (\ref{190906a}) to (\ref{190906d}) for the tidal dissipation.
We use a Runge-Kutta method of order 8 with an embedded error estimator of order 7 due to Dormand \& Prince, with step size control \citep{Hairer_etal_1993}.
The choice of the initial conditions is described in Sect.\,\ref{initcond}, and here we explore different values for the unknown parameters.

\subsection{Tidal dissipation ratio}
\llabel{tdratio}

Distinct tidal evolution behaviors depend on the ratio between the tidal dissipation in Charon and Pluto given by the $A$ parameter (Eq.\,(\ref{190911a})).
In a first set of simulations we therefore vary this parameter from $A=1$ to 16.

For a better comparison with the previous work by \citet{Cheng_etal_2014a}, we first adopt orbit \#1  (Table~\ref{TabIn}) for the primoridial system, which places Charon very close to Pluto in a not very eccentric orbit ($a/R_0 = 4$, $e=0.2$). 
We also assume that the initial obliquities of Pluto and Charon are very close to zero ($\theta_0 = \theta_1 = 0.001^\circ$).
For Charon, we assume an initial rotation period of 6~hours ($\w_1/n \approx 3$).
For Pluto, with expression (\ref{200526a}) we compute an initial rotation period of 3.7~hours ($\w_0/n \approx 4.5$), which is close to the centrifugal breakup limit (Sect.\,\ref{rotlimit}).

\begin{figure}
\centering
    \includegraphics[width=\columnwidth]{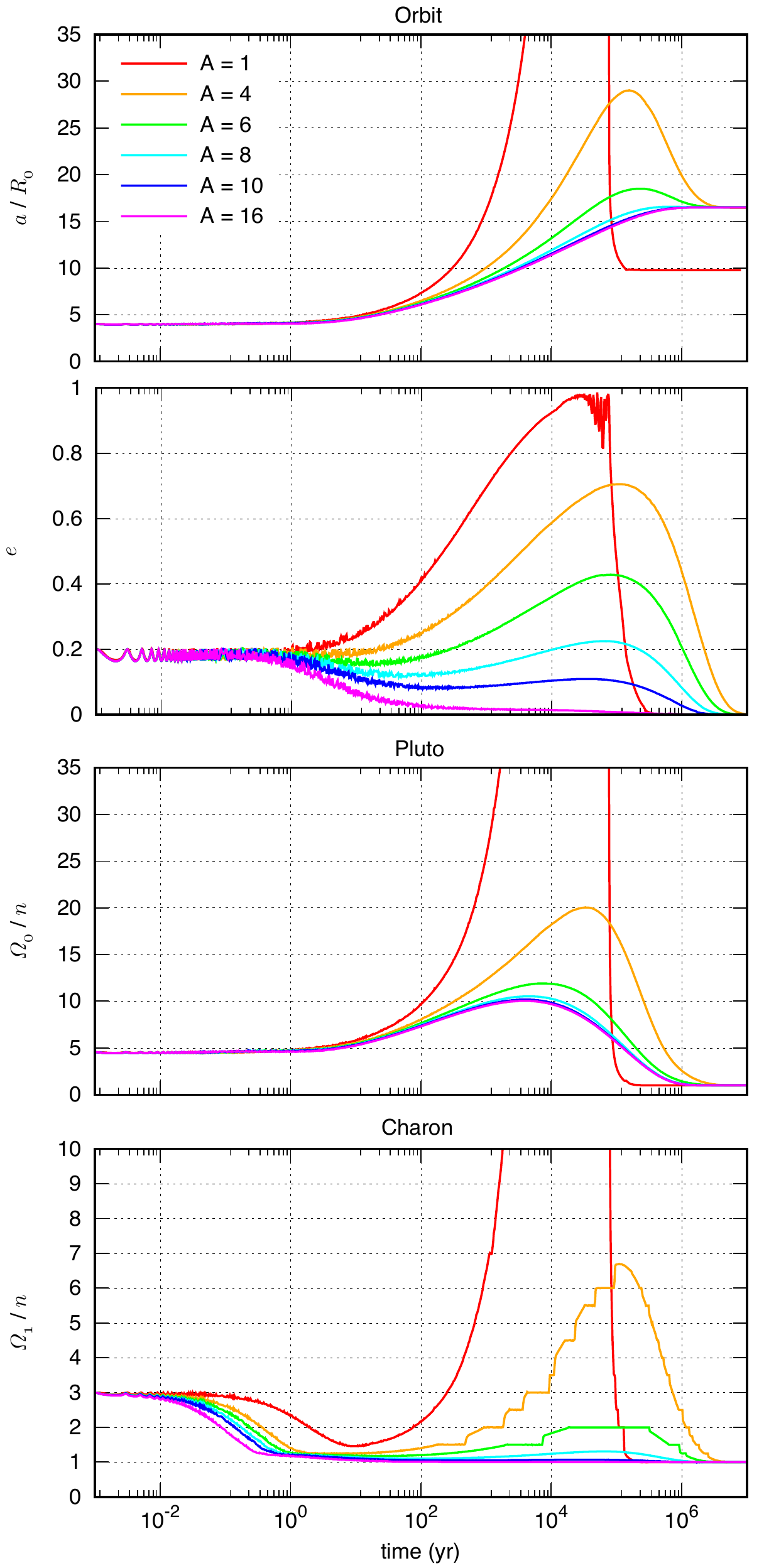}
 \caption{Tidal evolution of the Pluto--Charon binary for different values of the tidal parameter $A$, with initial $\theta_0 \approx 0^\circ$ and orbit \#1 (Table~\ref{TabIn}). We show the evolution of the semi-major axis and eccentricity (top), the rotation of Pluto (middle), and the rotation of Charon (bottom).} 
\label{fig_C_tides}
\end{figure}

\begin{figure}
\centering
    \includegraphics[width=\columnwidth]{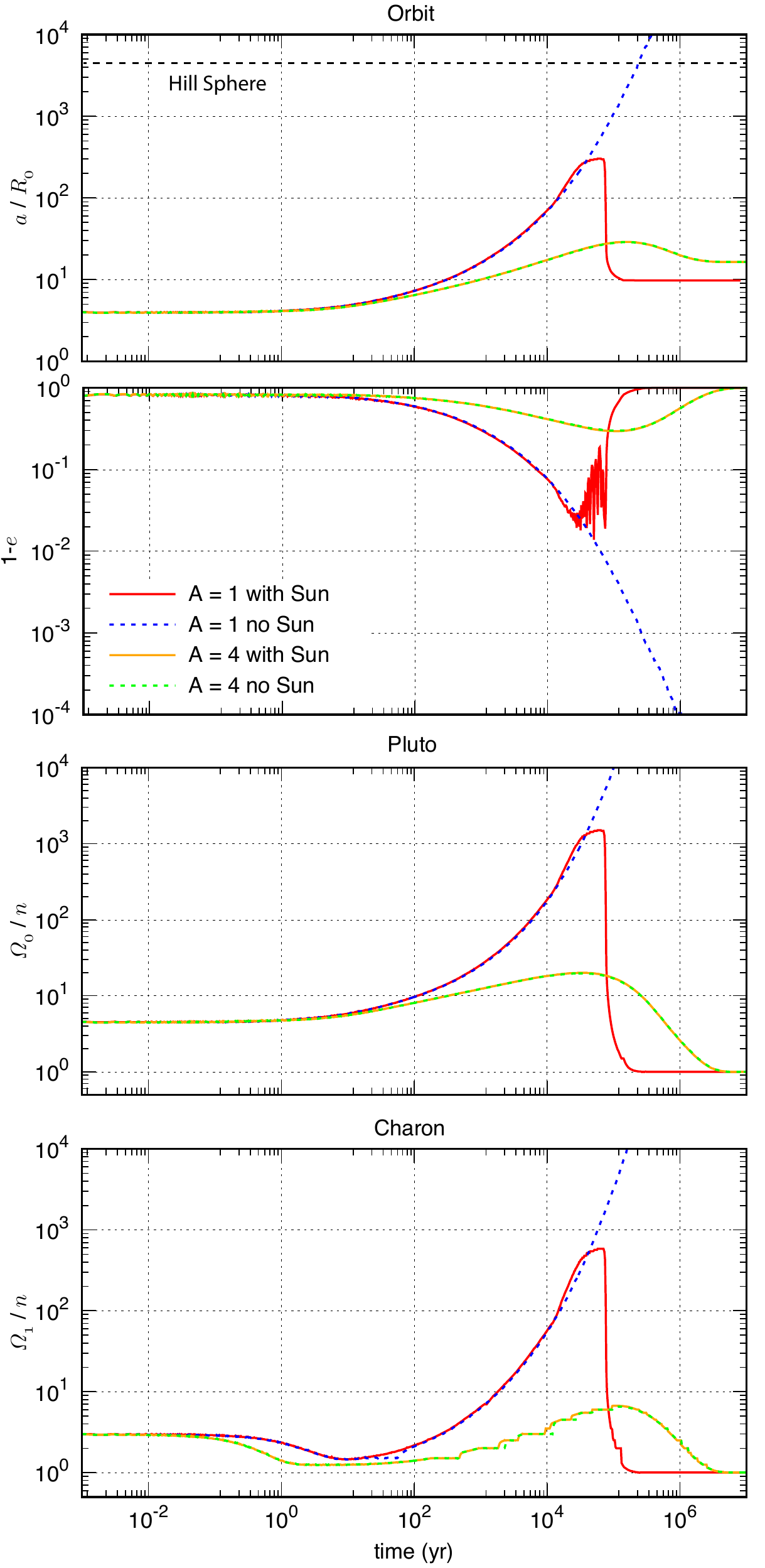}
 \caption{Tidal evolution of the Pluto--Charon binary for different values of the tidal parameter $A$, with initial $\theta_0 \approx 0^\circ$ and orbit \#1 (Table~\ref{TabIn}). We show the evolution obtained including the presence of the Sun (solid line) and without the presence of the Sun (dashed line). We plot the evolution of the semi-major axis and eccentricity (top), the rotation of Pluto (middle), and the rotation of Charon (bottom) on a log scale.} 
\label{fig_C_LK}
\end{figure}

Results are shown in Fig.\,\ref{fig_C_tides}.
As expected, since the initial obliquities are nearly zero, our results are in good agreement with those obtained by \citet{Cheng_etal_2014a}.
We observe that the full evolution takes less than $10$~Myr for all tidal ratios, 
the final semi-major axis always stabilizes at the present value $a/R_0=16.5$ (except for $A=1$, for which $a/R_0=11.4$),
 the final orbit is circularized, and
the rotation for both Pluto and Charon ends up captured in the synchronous resonance.

In the case of Charon, the spin evolution is extremely fast and
we observe multiple temporary captures in different spin-orbit resonances, in particular when the orbit is very eccentric.
These captures follow more or less the asymptotic equilibrium rotation given by expression (\ref{090520a}). 
Capture in spin-orbit resonances is a stochastic process, and so the output of our simulations for the rotation of Charon is only one among multiple possibilities.
However, when the eccentricity becomes higher or lower than a critical value, the resonances always become unstable and the rotation follows its course \citep{Correia_Laskar_2009, Correia_Laskar_2012}.

In the case of Pluto, the evolution can be understood with the averaged equations in Sect.\,\ref{evts}.
For large $A$ values ($A \gtrsim 10$), the eccentricity is damped more efficiently to zero because expression (\ref{200602c}) can be approximated by its last term (owing to Charon).
As a consequence, all $f_k(e) \approx 1$, and we get 
for the semi-major axis with $\theta_0 = 0$ (Eq.\,(\ref{200602b})):
\be
\frac{\dot a}{a} \approx  \frac{2 {\cal K}}{\beta a^2} \left(  \frac{\w_0}{n} - 1 \right)  \approx  - \frac{2 C_0}{\beta a^2} \frac{\dot \w_0}{n} 
\ . \llabel{200602a}
\ee
We therefore conclude that 
the semi-major axis always increases, until the rotation is synchronized with the orbit ($\w_0=n$).
Combining with expression (\ref{090515a}), we additionally get
\be
\frac{d}{dt} \left(\frac{\w_0}{n} \right) = \frac{\dot \w_0}{n} + \frac32 \frac{\w_0}{n} \frac{\dot a}{a} 
\approx \frac32 \left( \frac{\w_0}{n} - \frac{\beta a^2}{3 C_0} \right) \frac{\dot a}{a} 
\ . \llabel{200603a}
\ee
For the initial rotation and semi-major axis, we have $\w_0/n \approx 4.5$ and $\beta a^2 / 3 C_0 \approx 1.9$, and so the ratio $\w_0/n$ increases and moves away from synchronous equilibrium.
As the orbit expands, there is a turning point after which $\w_0/n$ decreases, for
\be
\frac{\w_0}{n} = \frac{\beta a^2}{3 C_0} \approx \left(\frac{a}{3 R_0}\right)^2 
\llabel{200603b} \ .
\ee
Using total angular momentum conservation (Eq.\,(\ref{190909b})), 
we have the additional constraint 
$ \w_0 \approx \beta \left(n_\obs a_\obs^2 - n a^2 \right) / C_0 $, 
which gives for the turning point 
\be
a / R_0 \approx 9.3 
\ , \quad \mathrm{and} \quad \w_0/n \approx 9.6 \ .
\ee
We note that the evolution observed for $\w_0/n$ is essentially due to the semi-major axis variation, because the rotation of Pluto slowly decreases.

For moderate $A$ values ($A \sim 7$), the eccentricity damping owing to tides raised on Charon is balanced by tides raised on Pluto, which tend to increase the eccentricity (first term in expression (\ref{200602c})).
As a result, the eccentricity initially remains approximately constant.
As the ratio $\w_0/n$ increases (Eq.\,(\ref{200603a})), the eccentricity slightly increases, but as soon as the ratio $\w_0/n$ decreases (Eq.\,(\ref{200603b})), the eccentricity is damped to zero.
The overall behavior is that the eccentricity only presents some small oscillations around its initial value throughout the evolution.
The semi-major axis and rotation rate evolution are similar to those for larger $A$ values.

For small $A$ values ($A \lesssim 4$), tides raised on Pluto control the orbital evolution (first term in Eqs.\,(\ref{200602b}) and (\ref{200602c})).
For initial rotations $\w_0/n \gg 1$, both the semi-major axis and the eccentricity rapidly increase to high values.
The semi-major axis reaches values much larger than the present value, while the eccentricity may attain values close to one.
In these extreme situations, the appocentre distance may approach the Hill sphere radius and the binary can become unstable \citep{Cheng_etal_2014a}.
Interestingly, this is not what we observe. 
At high eccentricities, there are angular momentum exchanges with the Sun because of Lidov-Kozai cycles that keep the eccentricity at values $e \lesssim 0.95$ and thus prevent the system from being destroyed (see Sect.\,\ref{LKc}).

\subsection{Lidov-Kozai cycles}
\llabel{LKc}

For small $A$ values, the semi-major axis and the eccentricity can grow to extremely high values. 
In Fig.\,\ref{fig_C_tides}, for $A=1$, the semi-major axis becomes so large at some point that it is not shown.
Therefore, in Fig.\,\ref{fig_C_LK}, we show again the tidal evolution for orbit \#1, but using a logarithmic scale. 
In addition, we also show the results of a simulation where the Sun is not included, that is, we integrate  only a two-body problem (Sect.\,\ref{pcbin}).

We observe that, for $A=4$, the evolution is similar for the simulations with and without the perturbations from the Sun.
However, for $A=1$, in the absence of the Sun the semi-major axis increases indefinitely and the eccentricity takes a value of almost one.
As a consequence, Charon would be lost.
We conclude that the stability of the Pluto--Charon binary can only be correctly addressed by taking into account a three-body problem.

\begin{figure}
\centering
    \includegraphics[width=\columnwidth]{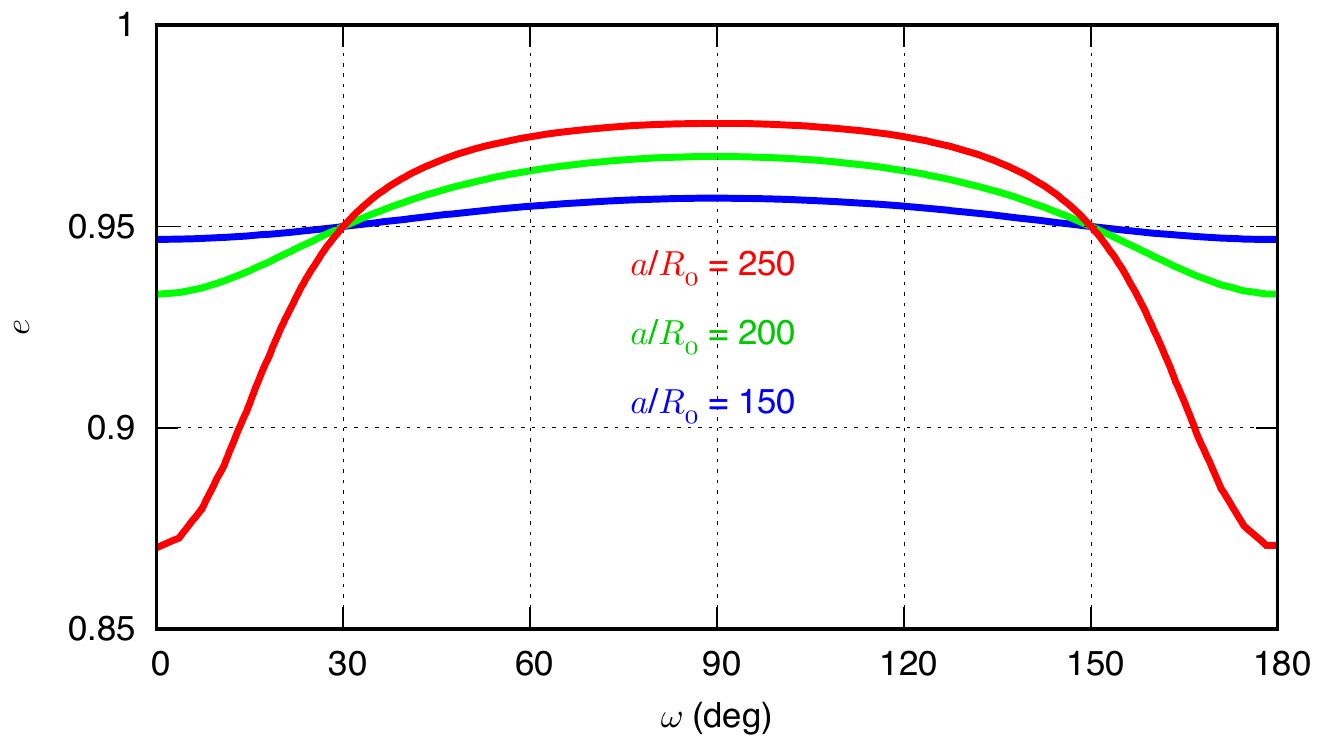}
 \caption{Level curves of the total energy (Eq.\,(\ref{090514a})) for different values of the semi-major axis, with $2 \pi / \w_0 = 8$~hours and high eccentricity. The initial condition for all curves is $\omega_\bin=30^\circ$, $e_\bin = 0.95$ and $i = 122^\circ$.} 
\label{lidov_kozai}
\end{figure}

Assuming zero obliquity for Pluto ($\theta_0=0$), and retaining only the main contributions, the total potential energy (Eq.\,(\ref{171124b})) can be simplified as \citep[e.g.,][]{Correia_etal_2013}
\begin{eqnarray}
U_T \!\!\!\! &= \!\!\!\! &  - \C_1 (1-e_\bin^2)^{-3/2} 
\nonumber \\
&& - \C_2 \Big[ (1 + \tfrac{3}{2} e_\bin^2) (1-\tfrac{3}{2} \sin^2 i) + \tfrac{15}{4} e_\bin^2 \sin^2 i \cos 2 \omega_\bin \Big]
\ , \llabel{090514a}
\end{eqnarray}
where $i$ is the mutual inclination, $\omega_\bin$ is the argument of the pericenter of the binary orbit measured from the line of nodes,  
\be
\C_1 = \frac{\kf \m_1 \w_0^2 R_0^5}{6 a_\bin^3}  \ , \quad \mathrm{and} \quad
\C_2 = \frac{\beta_\bin n_\sun^2 a_\bin^2}{4 (1-e_\sun^2)^{3/2}} 
 \ .  \llabel{110816a}
\ee
The term in $\C_1$ results from the oblateness of Pluto owing to rotation, while the term in $\C_2$ results from the quadrupole gravitational interactions with the Sun.
For close-in orbits, $\C_1 \gg \C_2$, and so the binary orbit precesses rapidly and the eccentricity remains approximately constant (in the absence of tides).
For $\C_1 \sim \C_2$, the two contributions are equivalent, which occurs for
\be
a_\bin / R_0 \sim \left(\w_0 / n_\sun \right)^{2/5} \sim 150 \ ,
\llabel{200911b}
\ee
with $2 \pi / \w_0 = 8$~hours.
At this stage, the gravitational interactions with the Sun become important in shaping the dynamics of the system.
In particular, the angular momentum of the binary can be transferred to the orbit of the Sun.
Indeed, because the two orbits have a mutual inclination of $i=122^\circ$, we can observe Lidov-Kozai cycles \citep{Lidov_1962, Kozai_1962}.
These latter consist in exchanges of eccentricity and mutual inclination, such that 
\be 
(1-e_\bin^2) \cos^2 i \approx const
\ . \llabel{101221d}
\ee
Replacing this condition in the expression of the total energy (Eq.\,(\ref{090514a})) gives us an integrable problem whose dynamics can be easily understood in terms of a $(e_\bin, \omega_\bin)$ diagram.
In Fig.\,\ref{lidov_kozai}, we show the level curves of the total energy for different values of the semi-major axis, with $2 \pi / \w_0 = 8$~hours and high eccentricity.
We observe that, for $a_\bin / R_0 = 150$ (Eq.\,(\ref{200911b})), there is only a small oscillation in the eccentricity.
However, as we increase the semi-major axis ($a_\bin / R_0 \ge 200$), the interactions with the Sun progressively reduce the minimum eccentricity.

When we use an average smaller eccentricity value in the evolution of the semi-major axis (Eq.\,(\ref{200602b})), we find that the semi-major axis slows down its expansion rate, and subsequently reverses its evolution. We therefore conclude that Lidov-Kozai cycles 
act as a protective mechanism.

\begin{figure}
\centering
    \includegraphics[width=\columnwidth]{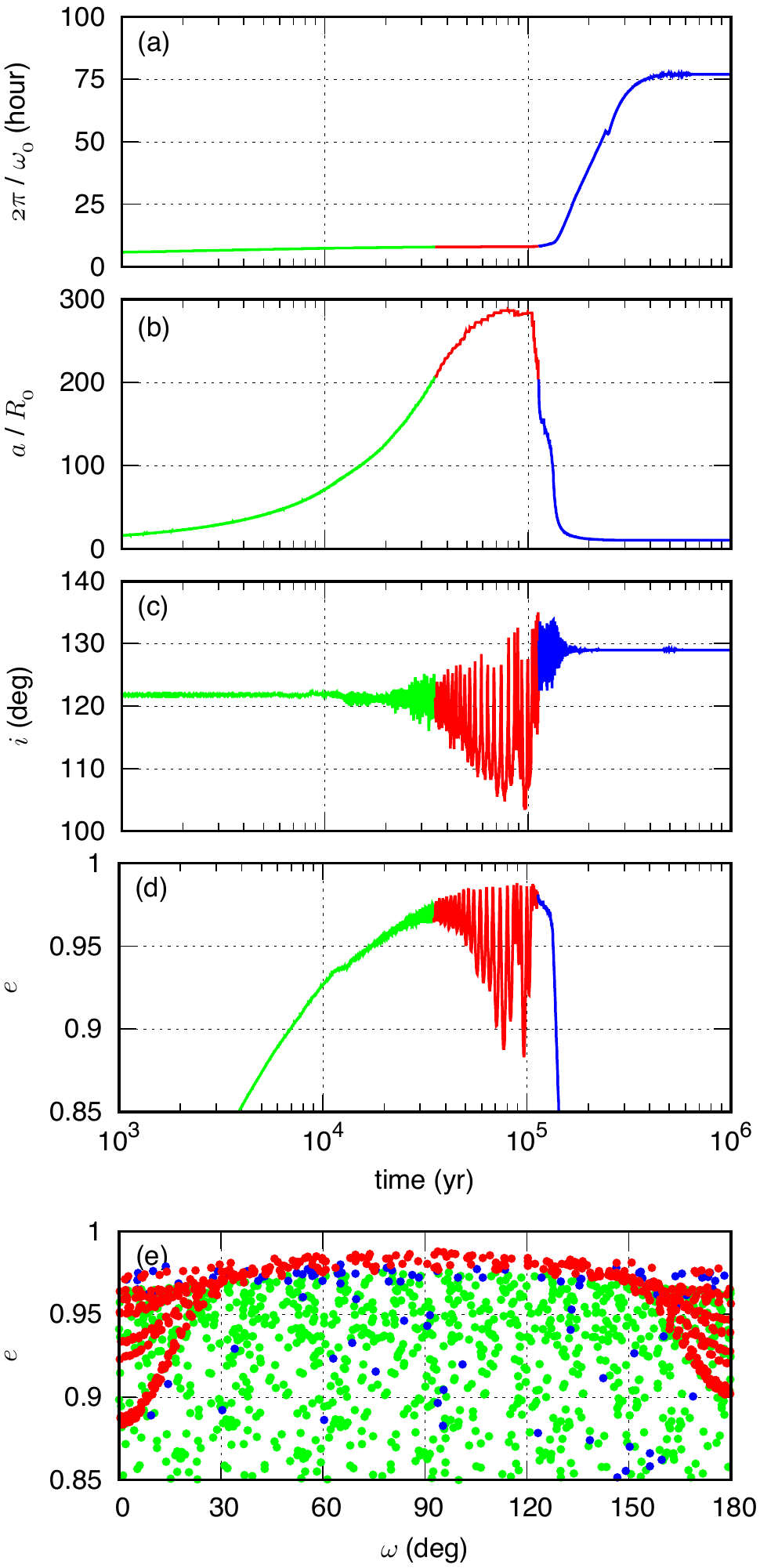}
 \caption{Tidal evolution of the Pluto--Charon binary for $A=1$, initial $\theta_0 \approx 0^\circ$ and orbit \#1 (Table~\ref{TabIn}). We plot the evolution of Pluto's rotation period (a), semi-major axis (b), inclination between orbital planes (c), and eccentricity (d) as a function of time, and the evolution in the $(e_\bin, \omega_\bin)$ diagram (e). The different colors highlight the three evolution stages, initial semi-major axis and eccentricity increase (green), Lidov-Kozai cycles (red), final semi-major axis and eccentricity damping (blue).} 
\label{fig_C_ew}
\end{figure}

In order to better understand this interesting scenario, in Fig.\,\ref{fig_C_ew} we show the critical stages of the evolution in detail for the orbit \#1 with $A=1$ (corresponding to the same evolution depicted in Figs.\,\ref{fig_C_tides} and \ref{fig_C_LK}).
In red color we highlight the evolution during the large semi-major axis and high eccentricity phase, where Lidov-Kozai cycles can occur.

Figure~\ref{fig_C_ew}\,(a) shows the evolution of Pluto's rotation period as a function of time.
We see that it is nearly constant and close to the initial value until the semi-major axis and the eccentricity drop to low values.
Figure~\ref{fig_C_ew}\,(b) shows the evolution of the semi-major axis for guidance.
Figures~\ref{fig_C_ew}\,(c) and (d) show the evolution of the mutual inclination and eccentricity as a function of time.
We observe significant oscillations of both parameters for large values of the semi-major axis ($a  / R_0 \ge 200 $, red color), corresponding to the Lidov-Kozai cycles.
According to expression (\ref{101221d}), when $\cos^2 i$ decreases (or increases), the eccentricity also decreases (or increases).
Finally, in Fig.\,\ref{fig_C_ew}\,(e), the evolution of the eccentricity is drawn in the diagram $(e_\bin, \omega_\bin)$.
We observe that during the large semi-major axis phase, the red points are distributed in agreement with the phase space shown in Fig.\,\ref{lidov_kozai}, confirming that the Lidov-Kozai oscillation is taking place.

The evolution shown in Fig.\,\ref{fig_C_ew} is not representative of the present system, because the final semi-major axis is below the present value.
However, it shows that it is theoretically
possible that the initial binary system was formed with a total angular momentum larger than what we observe today, provided that this excess was removed through the interactions with the Sun at periods of high eccentricity.

\subsection{Distinct initial orbits}

\begin{figure}
\centering
    \includegraphics[width=\columnwidth]{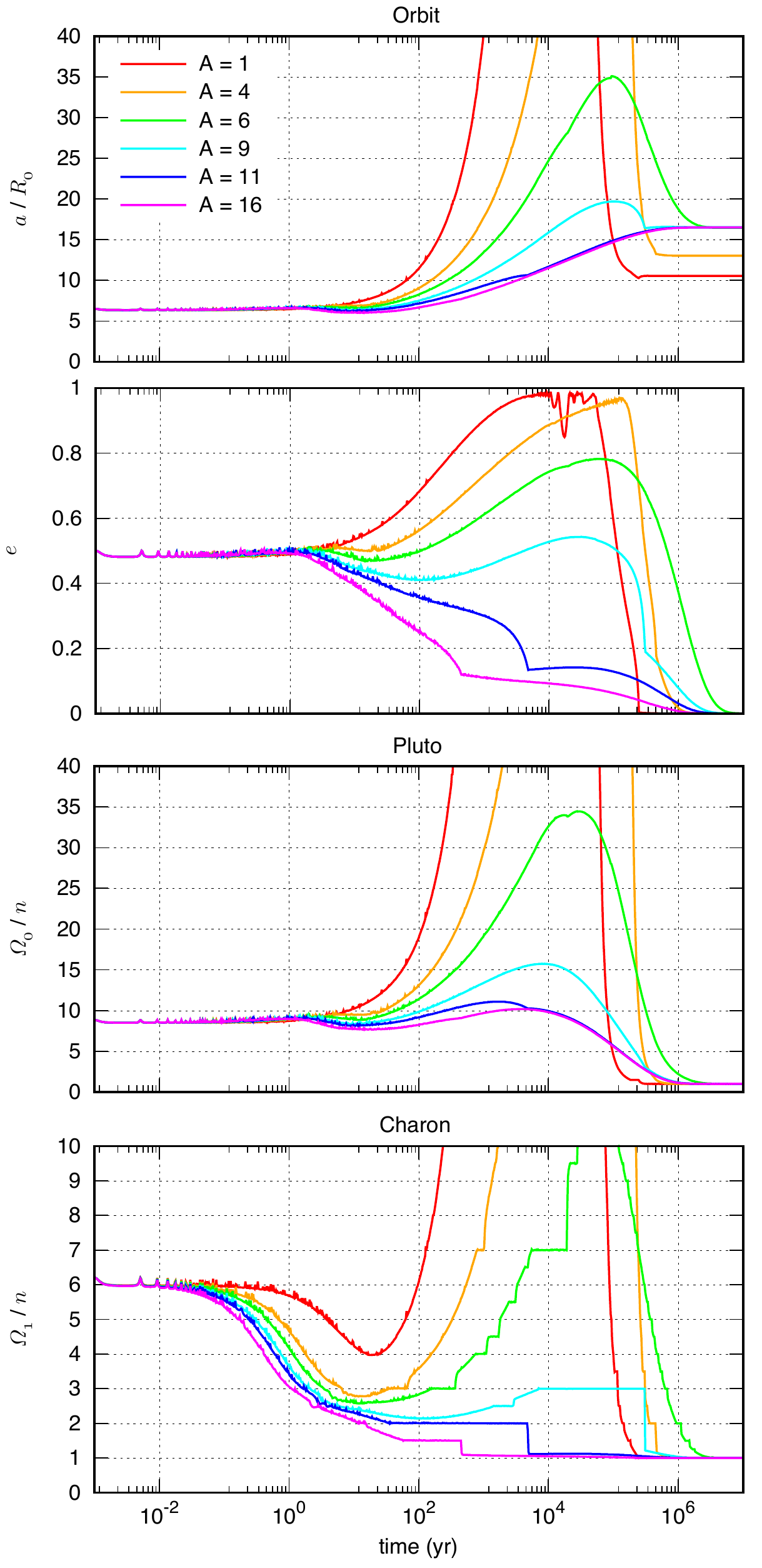}
 \caption{Tidal evolution of the Pluto--Charon binary for different values of the tidal parameter $A$, with initial $\theta_0 \approx 0^\circ$ and orbit \#2 (Table~\ref{TabIn}). 
} 
\label{fig_A_tides}
\end{figure}

\begin{figure}
\centering
    \includegraphics[width=\columnwidth]{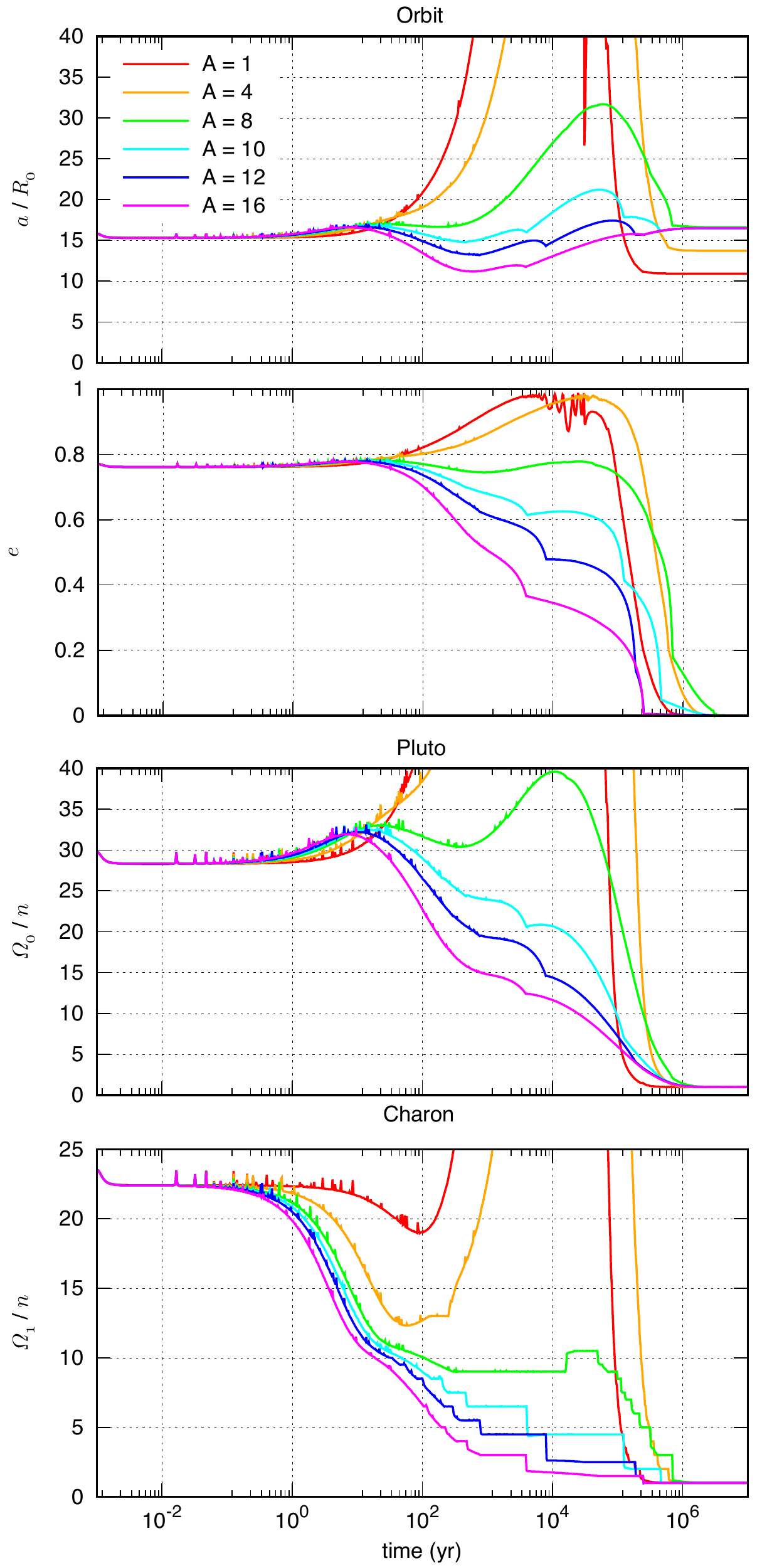}
 \caption{Tidal evolution of the Pluto--Charon binary for different values of the tidal parameter $A$, with initial $\theta_0 \approx 0^\circ$ and orbit \#3 (Table~\ref{TabIn}). 
 } 
\label{fig_B_tides}
\end{figure}

We repeat the same experiment from Sect.\,\ref{tdratio}, but adopting orbits \#2  and \#3 for the initial system (Table~\ref{TabIn}).
We assume again that the initial obliquities of Pluto and Charon are close to zero ($\theta_0 = \theta_1 = 0.001^\circ$), and 6~hours for the initial rotation period of Charon.
Due to the angular momentum conservation, Pluto's initial rotation period is now 4.1 and 4.7~hours for initial orbits \#2  and \#3, respectively (Eq.\,(\ref{190910zb})).
The tidal parameter $A$ is again varied from 1 to 16.

Orbit \#2 places Charon still close to Pluto, but in a moderately eccentric orbit ($a/R_0 = 6.5$, $e=0.5$), which gives $\w_0/n \approx 8.8$ and  $\w_1/n \approx 6.2$ for the initial rotation ratios.
On the other hand, orbit \#3 places Charon nearly at the present semi-major axis, but in a very eccentric orbit ($a/R_0 = 15.8$, $e=0.77$), which gives $\w_0/n \approx 29.8$ and  $\w_1/n \approx 23.5$.
Results for orbit \#2 are shown in Fig.\,\ref{fig_A_tides}, and results for orbit \#3 are shown in Fig.\,\ref{fig_B_tides}. 

We observe that the main features already present for the simulations with orbit \#1 persist:
(1) for large $A$ values the eccentricity is quickly damped to zero, while for small $A$ values it can increase to very high values; 
(2) the semi-major axis always evolves to the present value, except for small $A$ values;
(3) for small $A$ values Lidov-Kozai cycles occur and stabilize the system;
(4) the spin of Charon quickly evolves into spin-orbit resonances; and
(5) the rotation ratio of Pluto ($\w_0/n$) initially increases, then decreases towards the synchronous value.
Despite these global trends, the individual variations may present a number of subtle differences, either related to the fact that the initial orbit has a larger semi-major axis and a higher eccentricity, or to the spin evolution.

For instance, for orbit \#3, we observe that the semi-major axis remains more or less constant throughout the evolution, because the initial value is already close to the present value.
On the other hand, as the initial eccentricity is also higher, in order to keep the eccentricity nearly constant during most of the evolution, we need smaller $A$ values (weaker tides on Charon).

An interesting backreaction effect, previously unnoticed for orbit \#1, occurs between the spin of Charon and the orbital evolution.
For large $A$ values, for which tides raised on Charon dominate the evolution,
the semi-major axis and the eccentricity undergo sudden changes.
These striking modifications in the orbit result from a change in the spin of Charon, which switches to a different spin-orbit resonance (see Sect.\,\ref{sor}).

\subsection{Spin-orbit resonances}
\llabel{sor}

We consider a permanent residual deformation $J_2 = 6 \times 10^{-5}$ and $C_{22} = 10^{-5}$ in our model (Sect.\,\ref{shapeJC}), which allows spin-orbit resonances between the rotation rate and the mean motion
\citep[e.g.,][]{Colombo_1965, Goldreich_Peale_1966, Correia_Delisle_2019}.
These can be observed for both Pluto and Charon in all plots that show the ratio $\w/n$. 

The rotation of Charon quickly evolves into a slow rotation regime for which $\w_1 \sim n$, while the rotation of Pluto is much faster than the orbital period ($\w_0 \gg n$) most of the time.
As a consequence, resonant capture for Pluto is only observed towards the end of the evolution and often in the synchronous resonance, because at this point the eccentricity is already close to zero.
On the other hand, for Charon we observe a wide variety of resonant captures that follow the eccentricity evolution.

\begin{figure}
\centering
    \includegraphics[width=\columnwidth]{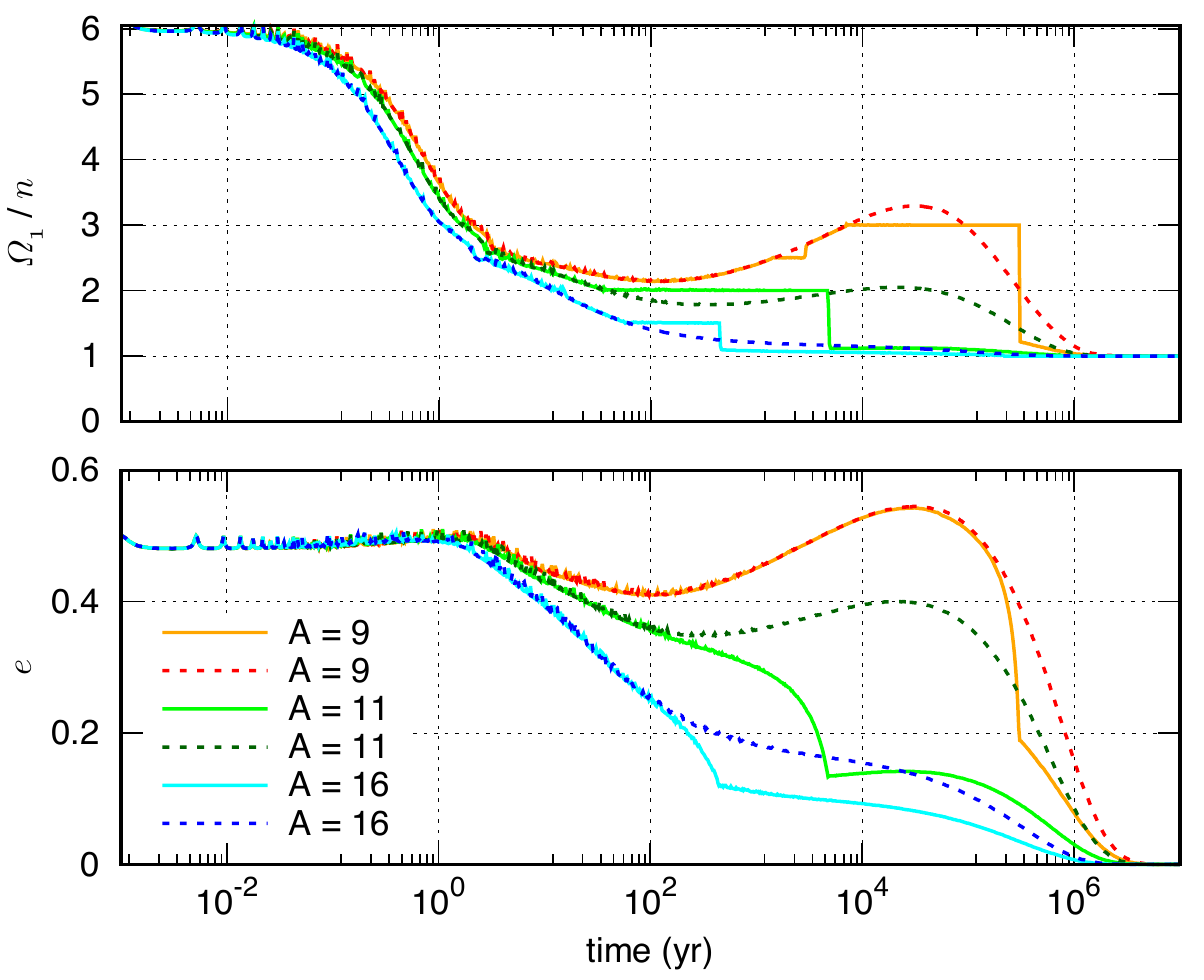}
 \caption{Tidal evolution of the Pluto--Charon binary for different values of the tidal parameter $A$ and permanent deformation, with initial $\theta_0 \approx 0^\circ$ and orbit \#2 (Table~\ref{TabIn}). We show the evolution of Charon's rotation rate (top) and eccentricity (bottom). Solid lines correspond to $J_2 = 6 \times 10^{-5}$ and $C_{22} = 10^{-5}$, and dashed lines to $J_2=C_{22}=0$. } 
\label{fig_A_pcap}
\end{figure}

Spin-orbit captures are interesting in the evolution of the Pluto--Charon binary, because they can modify the damping timescale of the eccentricity \citep{Cheng_etal_2014a}.
This backreaction effect was first described for exoplanets \citep{Rodriguez_etal_2012}, and it is particularly important when tides raised on Charon control the evolution.
Indeed, assuming that Charon is captured in the $\w_1/n = p$ resonance ($p$ is an half integer) with $\theta_1 = 0$, for large $A$ values (Eq.\,(\ref{090515c})) we have 
\be
\frac{\dot e}{e} \approx \frac{9 {\cal K} A}{\beta a^2} \Big( \frac{11}{18} f_4(e) p - f_5(e) \Big)
\ , \llabel{200624a}
\ee
which differs from what is obtained with $\w_1 = \w_{\rm e}$ (Eq.\,(\ref{200602c})):
\be
\frac{\dot e}{e} \approx -\frac{7 {\cal K} A}{2 \beta a^2} f_6(e) (1-e^2) 
\ . \llabel{200624a}
\ee

In Fig.\,\ref{fig_A_pcap}, we show some examples for Charon's rotation and eccentricity evolution with and without the permanent deformation.
We adopt initial orbit \#2, Charon's initial rotation period of 6~hours, $\theta_0 = \theta_1 = 0.001^\circ$, and $A=9$, 11, and 16 (as in Fig.\,\ref{fig_A_tides}).
In the absence of permanent deformation (dashed lines), the rotation of Charon follows the exact equilibrium value given by expression (\ref{090520a}).
In the case with $C_{22} = 10^{-5}$ (solid lines), capture in spin-orbit resonances always occurs at some point.
We observe that, just after capture in resonance, the eccentricity evolution is modified.
On the other hand, once a spin-orbit resonance is destabilised, the eccentricity evolution immediately slows down.
In general, capture in spin-orbit resonances tend to damp the eccentricity more efficiently.
At the end of the evolution, the rotation always ends captured in the synchronous resonance and the eccentricity is damped to zero.

\subsection{Initial obliquity of Pluto}
\llabel{initobl}

The large collision that gave rise to the Pluto--Charon binary likely produced a misaligned system, that is, the orbital plane of Charon and the equatorial plane of Pluto were tilted \citep{Canup_2005, Canup_2011}.
Therefore, in this section we test the consequences of different initial obliquities for Pluto, $\theta_0$.
For Charon we always assume an initial rotation period of 6~hours and $\theta_1=0.001^\circ$.
We fix the tidal dissipation ratio at $A=4$, and run a set of numerical simulations for two initial orbital configurations (Table~\ref{TabIn}).

\begin{figure*}
\centering
    \includegraphics[width=\textwidth]{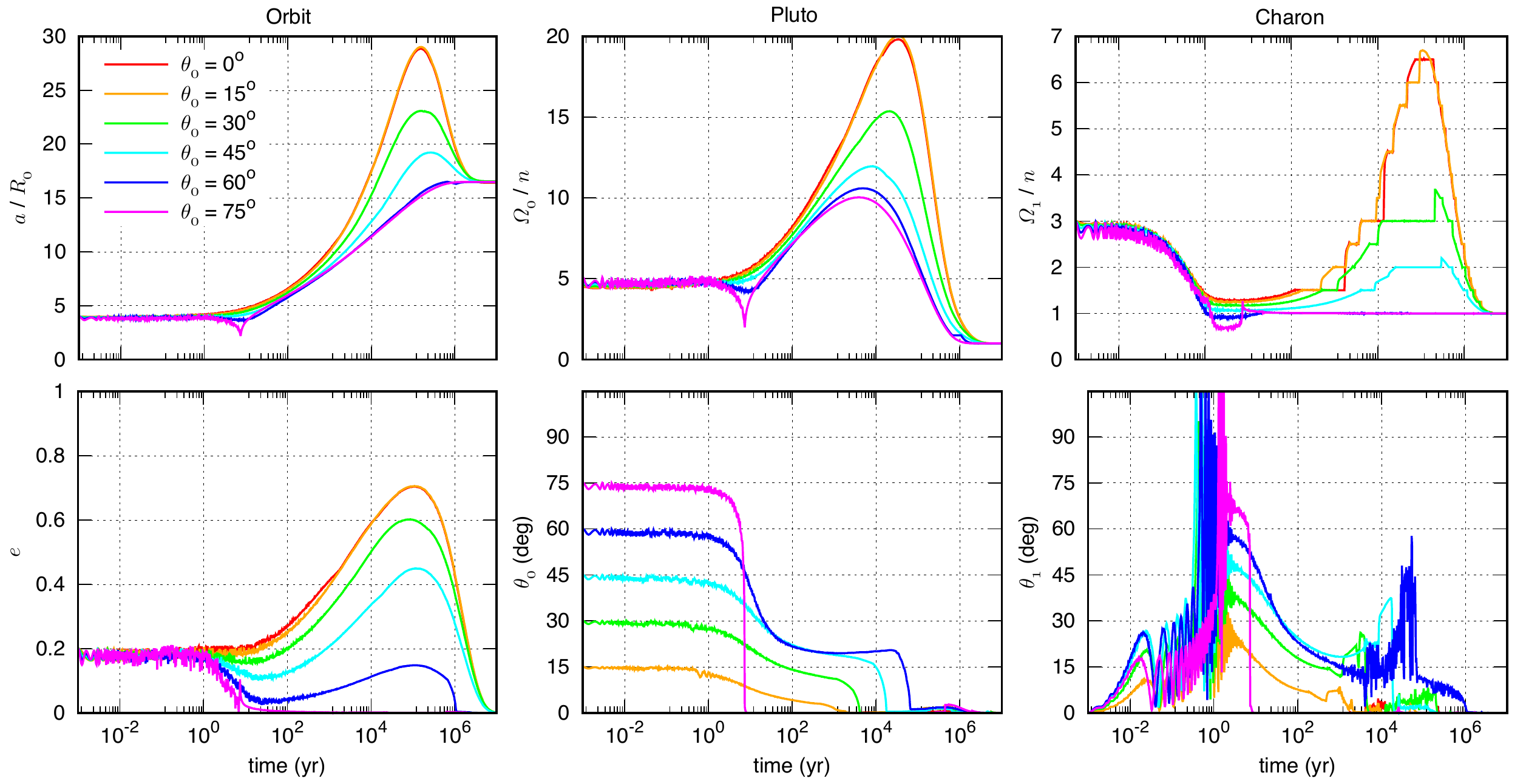}
 \caption{Tidal evolution of the Pluto--Charon binary for different values of Pluto's initial obliquity, $\theta_0$, with $A=4$ and initial orbit \#1 (Table~\ref{TabIn}). 
} 
\label{fig_C_obl}
\end{figure*}

Figure~\ref{fig_C_obl} shows the results for orbit \#1.
As expected, regardless of the initial obliquity, in all simulations the orbit is circularised, the final rotation for both Pluto and Charon ends captured in the synchronous resonance, and the obliquities end very close to zero, which corresponds to the final outcome of tidal evolution \citep{Hut_1980, Adams_Bloch_2015}.
Nonetheless, for all simulations we observe an interesting early excitation of Charon's obliquity. 
Pluto is initially very oblate due to fast rotation and its equator is not aligned with the initial orbit of Charon ($\theta_0 > 0^\circ$). 
As a consequence, the gravitational torque of Pluto on Charon induces large obliquity variations, even in the absence of tides.
These variations show that the initial choice for the spin of Charon is irrelevant, because its evolution is rapidly controlled by the external torques.

We observe that for initial obliquities $\theta_0 \lesssim 30^\circ$, the orbital and spin evolution are very similar.
This is particularly true for the runs with $\theta_0 = 0^\circ$ and $\theta_0 = 15^\circ$, for which we can only detect changes in the obliquity evolution and in the distribution of Charon's rotation capture in spin-orbit resonances.
Nevertheless, as we increase the initial obliquity of Pluto, a striking difference arises: although we adopt $A=4$ in all runs, simulations with higher initial obliquity values resemble those with lower initial obliquities, but with larger $A$ values (see Fig.\,\ref{fig_C_tides}).

For $A=4$, the orbital evolution is mainly controlled by tides raised on Pluto, for which the semi-major axis and the eccentricity can grow to high values when the initial obliquity is low (Sect.\,\ref{tdratio}).
However, we observe that, as we increase the initial obliquity, the semi-major axis and the eccentricity initially decrease.
Indeed, for obliquity values close to $90^\circ$, we get $\cos \theta_0 \approx 0$, and expressions (\ref{200602b}) and (\ref{200602c}) become always negative.
After the obliquity is damped, the semi-major axis and the eccentricity can increase again, but now they restart from a lower value.
As a consequence, they cannot reach values as high as in the case of an initial low obliquity. 

Another interesting feature is that we cannot choose an arbitrary high initial obliquity.
One reason for this is that the initial rotation rate required to conserve the total angular momentum of the system increases with the obliquity (Eq.\,(\ref{200526a})) and may exceed the rotational breakup limit.
However, the main restriction is related to the semi-major axis evolution.
We have just seen that high initial obliquities, in particular those with $\theta_0 \ge 90^\circ$, initially reduce the semi-major axis.
Thus, the semi-major axis may become so small that the two bodies collide.
In the simulation with orbit \#1 and $A=4$, this limit is around $\theta_0 = 77^\circ$
(for $\theta_0 = 75^\circ$ the minimum periapse distance is already $1.9\,R_0$, very close to a physical collision).
Since the magnitude of tidal effects is proportional to $a^{-6}$, 
initial obliquities close to the limit value are also damped to zero more efficiently (Eq.\,(\ref{090520d})).

\begin{figure*}
\centering
    \includegraphics[width=\textwidth]{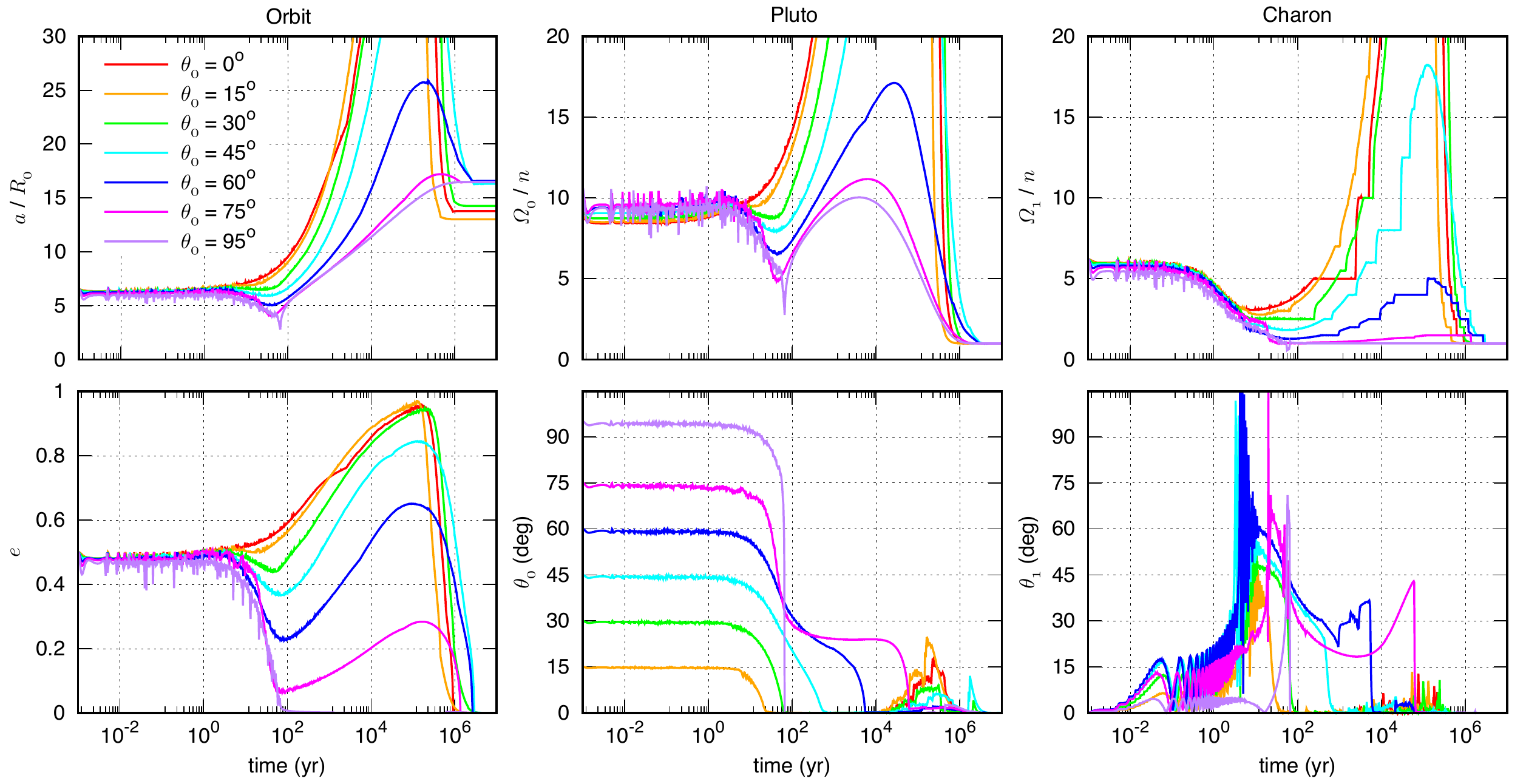}
 \caption{Tidal evolution of the Pluto--Charon binary for different values of Pluto's initial obliquity, $\theta_0$, with $A=4$ and initial orbit \#2 (Table~\ref{TabIn}). 
} 
\label{fig_A_obl}
\end{figure*}

Figure\,\ref{fig_A_obl} shows the results for initial orbit \#2.
We observe the same behavior already described for the simulations with orbit \#1, 
but even more pronounced.
As this orbit starts with a larger semi-major axis of $6.5\,R_0$, the limit initial obliquity that prevents a collision between Pluto and Charon is higher.
We estimate this limit at $\theta_0 \approx  96^\circ$. 
For initial orbit \#2 and $A=4$, initially low obliquities resulted in very eccentric orbits, for which the system could only be stabilized owing to Lidov-Kozai cycles (Sect.\,\ref{LKc}).
As we increase the initial obliquity, the eccentricity is initially damped and no longer grows to extreme values; exchanges of angular momentum with the Sun no longer occur and therefore the final configuration corresponds to the presently observed system.

\bfx{We conclude that the initial obliquity is a key variable to take into account in the past history of the Pluto--Charon system;
it is as important as the tidal parameter $A$, because it acts on the eccentricity in a similar way.
Indeed, in order to keep the eccentricity small throughout the evolution, it is no longer required that tides raised on Charon dominate the orbital evolution;
it is enough that Pluto starts with a high initial obliquity.}

\section{Conclusion}
\llabel{concdisc}

In this paper we revisit the tidal evolution of the Pluto--Charon binary.
We follow the system from its formation 
until the present-day configuration.
We considered a 3D model for the orbits and spins, permanent triaxial deformations, and the presence of the Sun.
All these effects prove to be important and modify the evolution of the system under certain conditions.

Previous 2D studies revealed that the orbital evolution of the Pluto--Charon system is essentially controlled by the ratio between tides raised on Charon and Pluto, which is modeled in our study by the $A$ tidal parameter (Eq.\,(\ref{190911a})).
In order to prevent the eccentricity from rising to extremely high values, 2D studies need to adopt large $A$ values, which means that tides raised on Charon dominate the evolution.
However, we observed that for high initial obliquities of Pluto ($\theta_0 \gtrsim 60^\circ$), the eccentricity is damped efficiently even for small $A$ values.
As a consequence, the eccentricity may preserve a nonzero and not overly high value during most of the evolution, even when tides raised on Pluto dominate the evolution.

Another possible way to damp the eccentricity more efficiently is through capture in spin-orbit resonances.
The rotation of Charon quickly evolves into a slow rotation regime ($\w_1 \sim n$), where capture is possible.
We do observe a wide variety of resonant captures, which are enhanced for eccentric orbits.
In turn, the eccentricity evolution also depends on the rotation rate of Charon, and so when a capture occurs, we have a backreaction effect in the orbit that modifies its evolution.
This effect was first described for exoplanets \citep{Rodriguez_etal_2012}, but it is also important in the Pluto--Charon system.

For small $A$ values and low initial obliquity for Pluto, the semi-major axis and the eccentricity can grow to high values.
Charon could then approach the Hill sphere radius and escape.
Instead, we observe that when the eccentricity is close to 0.95, there are angular momentum exchanges with the Sun through Lidov-Kozai cycles that help the Pluto--Charon binary to remain bounded.
It is only possible to observe this interesting scenario with a 3D model, because the mutual inclination between the orbit of the binary and the orbit of the Sun is around $122^\circ$.
The stability of the four small satellites around the Pluto--Charon binary is difficult to explain within this scenario if they were already present prior to the tidal expansion of Charon's orbit \citep[e.g.,][]{Smullen_Kratter_2017, Woo_Lee_2018}.
However, it has been shown that these satellites may have formed after the system settled into its present configuration through an impact on Charon \citep{Bromley_Kenyon_2020}.

Here we adopted a viscous linear model for tides.
Tidal evolution in the Pluto--Charon binary occurs within the first 10~Myr after formation, when the two bodies are likely still mostly melt and fluid, and so this model seems appropriate.
This model also provides simple expressions for the tidal evolution that allow us to interpret the output of the numerical simulations more easily.
Different and eventually more realistic tidal models could be attempted in future studies \citep[e.g.,][]{Renaud_etal_2020}.
These could modify the evolution timescales and the capture probabilities in spin-orbit resonances, but the main conclusions from this study, enumerated above, should remain valid.

The model described in Sect.\,\ref{model} of this paper is the most complete model implemented so far for the study of the tidal evolution of the Pluto--Charon system. 
It is presented here in a very general formulation, and so it can be easily extended to the study of any binary system perturbed by an external body.
A straightforward application is the Earth--Moon system, although in this case the evolution timescale is much longer.
In order to improve our model for tidal evolution of binary systems, future work should also include planetary perturbations \citep[e.g.,][]{Correia_Laskar_2001, Correia_Laskar_2004}.

\begin{acknowledgements}
We thank T. Boekholt and D. Carvalho for discussions. 
This work was supported by
CFisUC (UIDB/04564/2020 and UIDP/04564/2020),
PHOBOS (POCI-01-0145-FEDER-029932), and
ENGAGE SKA (POCI-01-0145-FEDER-022217),
funded by COMPETE 2020 and FCT, Portugal.
\end{acknowledgements}

\bibliographystyle{aa}
\bibliography{\bibpath correia}

\end{document}